\documentclass[twocolumn,times]{aastex61}
\pdfoutput=1 
\usepackage{amsmath,amstext,natbib,float,epsfig,multirow}
\usepackage[figure,figure*]{hypcap}

\newcommand{\lyaf}[1]{Ly$\alpha$ forest}
\newcommand{\lya}[1]{Ly$\alpha$}
\newcommand{\kms}{\ensuremath{\mathrm{km\;s^{-1}}}}
\newcommand{\hMpc}{\ensuremath{h^{-1}\,\mathrm{Mpc}}}

\newcommand{\hMpccube}{\ensuremath{\, h^{-3}\,\mathrm{Mpc}^3 }}

\newcommand{\dperp}{\ensuremath{\langle d_{\perp} \rangle}}

\newcommand{\ang}{\ensuremath{\mathrm{\AA}}}

\newcommand{\waveion}[3]{\ion{#1}{#2} $\lambda$#3}

\newcommand{\zbg}{\ensuremath{z_\mathrm{bg}}}
\newcommand{\za}{\ensuremath{z_\alpha}}

\newcommand{\cmd}{\mathbf{C}_\mathrm{MD}}
\newcommand{\cdd}{\mathbf{C}_\mathrm{DD}}

\newcommand{\deltarec}{\ensuremath{\delta_F^{\mathrm{rec}}}}

\shorttitle{CLAMATO IGM Tomography DR2}
\shortauthors{Horowitz et. al.}

\begin{document}

\title{Second Data Release of the COSMOS Lyman-Alpha Mapping And Tomography Observations:\\ 
The First 3D Maps of the Detailed Cosmic Web 
at $2.05 < \lowercase{z} < 2.55$}
\author[0000-0001-7832-5372]{Benjamin Horowitz}

\affiliation{Department of Astrophysical Sciences, Princeton University, Princeton, NJ 08544, USA}
\affiliation{Lawrence Berkeley National Laboratory, 1 Cyclotron Road, Berkeley, CA 94720, USA}

\author[0000-0001-9299-5719]{Khee-Gan Lee}
\affiliation{Kavli IPMU (WPI), UTIAS, The University of Tokyo, Kashiwa, Chiba 277-8583, Japan}

\author[0000-0002-5934-9018]{Metin Ata}
\affiliation{Kavli IPMU (WPI), UTIAS, The University of Tokyo, Kashiwa, Chiba 277-8583, Japan}

\author[0000-0002-2003-4465]{Thomas M\"uller}
\affiliation{Max Planck Institute for Astronomy, K\"{o}nigstuhl 17, D-69117 Heidelberg, Germany}

\author{Alex Krolewski}
\affiliation{AMTD Fellow, Waterloo Centre for Astrophysics, University of Waterloo, Waterloo ON N2L 3G1, Canada}
\affiliation{Perimeter Institute for Theoretical Physics, 31 Caroline St. North, Waterloo, ON NL2 2Y5, Canada}

\author[0000-0002-7738-6875]{J.\ Xavier Prochaska}
\affiliation{Department of Astronomy and Astrophysics, University of California at Santa Cruz, 1156 High Street, Santa Cruz, CA 95064, USA}
\affiliation{University of California Observatories, Lick Observatory, 1156 High Street, Santa Cruz, CA 95064, USA}
\affiliation{Kavli IPMU (WPI), UTIAS, The University of Tokyo, Kashiwa, Chiba 277-8583, Japan}

\author{Joseph F.\ Hennawi}
\affiliation{Department of Physics, Broida Hall, University of California at Santa Barbara, Santa Barbara, CA 93106, USA}

\author{Martin White}
\affiliation{Department of Astronomy, University of California at Berkeley, New Campbell Hall,
Berkeley, CA 94720, USA}
\affiliation{Lawrence Berkeley National Laboratory, 1 Cyclotron Road, Berkeley, CA 94720, USA}

\author{David Schlegel}
\affiliation{Lawrence Berkeley National Laboratory, 1 Cyclotron Road, Berkeley, CA 94720, USA}

\author{R.\ Michael Rich}
\affiliation{Department of Physics and Astronomy, University of California at Los Angeles, Los Angeles, CA 90095, USA}

\author{Peter E.\ Nugent}
\affiliation{Lawrence Berkeley National Laboratory, 1 Cyclotron Road, Berkeley, CA 94720, USA}
\affiliation{Department of Astronomy, University of California at Berkeley, New Campbell Hall,
Berkeley, CA 94720, USA}

\author{Nao Suzuki}
\affiliation{Kavli IPMU (WPI), UTIAS, The University of Tokyo, Kashiwa, Chiba 277-8583, Japan}

\author{Daichi Kashino}
\affiliation{Institute for Advanced Research, Nagoya University, Furocho, Chikusa-ku, Nagoya, 464-8601, Japan}

\author[0000-0002-6610-2048]{Anton M. Koekemoer}
\affiliation{Space Telescope Science Institute, 3700 San Martin Dr., Baltimore, MD 21218, USA} 

\author[0000-0002-1428-7036]{Brian C. Lemaux}
\affiliation{Gemini Observatory, NSF’s NOIRLab, 670 N. A’ohoku Place, Hilo, Hawai’i, 96720, USA}
\affiliation{Department of Physics and Astronomy, University of California, Davis, One Shields Ave., Davis, CA 95616, USA}

\correspondingauthor{Benjamin Horowitz}
\email{bhorowitz@princeton.edu}

\begin{abstract}
We present the second data release of the COSMOS Lyman-Alpha Mapping And Tomography Observations (CLAMATO) Survey conducted with the LRIS spectrograph on the Keck-I telescope. This project used \lyaf{} absorption in the spectra of faint star forming galaxies and quasars at $z \sim 2-3$ to trace neutral hydrogen in the intergalactic medium. In particular, we use 320 objects over a footprint of $\sim 0.2$ deg$^2$ to reconstruct the absorption field at $2.05 < z < 2.55$ at $\sim 2$ \hMpc resolution. We apply a Wiener filtering technique to the observed data to reconstruct three dimensional maps of the field over a volume of $4.1 \times 10^5$  \hMpccube. In addition to the filtered flux maps, for the first time we infer the underlying dark matter field through a forward modeling framework from a joint likelihood of galaxy and \lyaf{} data, finding clear examples of the detailed cosmic web consisting of cosmic voids, sheets, filaments, and nodes. In addition to traditional figures, we present a number of interactive three dimensional models to allow exploration of the data and qualitative comparisons to known galaxy surveys. We find that our inferred over-densities are consistent with those found from galaxy fields. Our reduced spectra, extracted Lyman-$\alpha$ forest pixel data, and reconstructed tomographic maps are available publicly at \url{https://doi.org/10.5281/zenodo.7524313}.
\end{abstract}

\section{Introduction}

The \lyaf{} is a a collection of absorption features present in galaxy and quasar (QSO) spectra caused by the presence of neutral hydrogen, HI, in the diffuse intergalactic medium (IGM) along the line of sight \citep{1965ApJ...142.1633G}. By interpreting the neutral hydrogen as a biased tracer of the underlying mass distribution, 
the \lyaf{} has emerged over the past few decades as a key probe of large scale structure for cosmological analysis
 \citep[e.g.,][]{croft:1998,mcdonald:2006, slosar:2011, 
busca:2013}. At high redshifts, $z \gtrsim 2.0$, galaxy spectroscopic surveys become increasingly expensive while the IGM is still has an appreciable neutral fraction \citep{2016ARA&A..54..313M} and the rest frame \lya\\  ($\lambda = 1215.67 \r{A}$) redshifts into the optical atmospheric window, making this an optimal redshift for \lyaf{} observations. 

Existing studies of the \lyaf{} have focused on QSOs since they are the brightest ultraviolet sources at this redshift. Due to their comparatively low number densities, studies with QSOs have been primarily focused on one dimensional line of sight analysis \citep[e.g.][]{2005PhRvD..71f3534V} or, in the case of the BOSS \citep{eisenstein:2011, dawson:2013} and eBOSS surveys \citep{2021PhRvD.103h3533A} on correlations at scales of $\dperp \sim 20\,\hMpc$. These separations are insufficient for resolving individual cosmic structures in the transverse direction, but more than sufficient for the surveys main goal of measuring the baryon acoustic oscillation
signal in the \lyaf{} \citep{busca:2013, slosar:2013, kirkby:2013,
 font-ribera:2014a,delubac:2015, bautista:2017, du-mas-des-bourboux:2017}.

Going to significantly fainter QSO sources only leads to marginal improvement in terms of target number densities and average line of sight separation due to the shallow slope of the QSO luminosity function \citep{2008ApJ...675...49S,palanque-delabrouille:2013,2016A&A...587A..41P}.
The number of potential lines of sight can increase dramatically by expanding surveys to include targeting UV-emitting star-forming galaxies at $z > 2$, known as ``Lyman Break Galaxies" \citep[LBGs;][]{steidel:1996}. It was shown in \citet{lee:2014} that a $g = 24.5$ survey magnitude limit including LBGs would lead to $\sim 1500$ deg$^{-2}$ density of lines of sight  , corresponding to $\dperp  \sim 2.5\,\hMpc$ at $z \sim 2.3$ vs. $\dperp  \sim 7.5\,\hMpc$ using only QSOs. The properties of individual LBG spectra were further characterized in \citet{2020AJ....160...37M}.

With the increased density of lines of sight  available, it becomes possible to create three dimensional tomographic reconstructions of the underlying cosmic web. This approach was first discussed in \citet{pichon:2001} and \citet{caucci:2008}, while the first pilot demonstration of tomographic \lyaf{} observations were done in 
\citet{lee:2014a}. These early observations were later expanded in \citet{lee:2016} and the first data release of the COSMOS Lyman Alpha Mapping and Tomography Observations (CLAMATO) survey \citep{2018CLAMATO}. Results from the CLAMATO survey have been used to identify and catalogue the highest redshift cosmic voids \citep{2018ApJ...861...60K}. Along with these observations, there have been significant theoretical developments in using tomographic observations for identification and analysis of high redshift protocluster \citep{stark:2015a,stark:2015}, voids \citep{2018ApJ...861...60K}, and cosmic web geometry \citep{lee:2016a}. 

Since the release of the first CLAMATO results, the same Wiener filter reconstruction technique was applied to Stripe 82 region of the eBOSS survey, creating a coarse tomographic map with $\dperp  \sim 13.0\,\hMpc$ over a large cosmic volume \citep{2020JCAP...07..010R}. The ongoing \lya{} Tomography IMACS Survey (LATIS) survey will reach a comparable sight-line density as CLAMATO over a subset of the COSMOS and CFHTLS fields \citep{2020ApJ...891..147N}. The upcoming Subaru Prime Focus Spectrograph \citep{sugai:2015} will have a significant IGM tomography component, which will enable 3D reconstructions over a substantial sky fraction ($\sim 15$ deg$^2$), representing an order of magnitude increase over LATIS survey. 
In the future, there are proposed tomography programs for the Maunakea Spectroscopic Explorer \citep{mcconnachie:2016}, Thirty Meter Telescope Wide-Field Optical Spectrograph \citep{skidmore:2015}, and the European Extremely Large  Telescope Multi-Object Spectrograph \citep{hammer:2016}, which would enable tomography at even higher special resolutions \citep{2019A&A...632A..94J} as well as surveys like WEAVE-QSO on the William Herschel Telescope \citep{2016sf2a.conf..259P} which will cover significant sky area (over $~400$ deg$^2$) abet at lower spatial resolution \citep{2022MNRAS.514.1359K}.

Along with the ongoing and proposed tomographic observational programs, there is significant theoretical interest in creating more accurate maps of the reconstructed \lyaf{} flux \citep{2021arXiv210212306L} and underlying dark matter density. Dynamic forward modeling approaches can push reconstructions to smaller scales than the standard Wiener filter method \citep{2019ApJ...887...61H} as well as allow joint reconstructions with overlapping galaxy observations \citep{2021ApJ...906..110H}. Additional science cases recently explored for IGM tomography include constraining galactic feedback models \citep{2020arXiv200714253N}, mapping quasar light echos \citep{2019ApJ...882..165S}, and studying variations in the UV background \citep{2019MNRAS.487.5346T,2021ApJ...909..117M}.

In this Article, we present the second public data release of the COSMOS Lyman-Alpha Mapping And 
Tomographic Observations (CLAMATO) survey\footnote{Website: \url{http://clamato.lbl.gov}}. These observations were taken on the LRIS spectrograph \citep{oke:1995,steidel:2004} on the Keck-I telescope near the summit of Mauna Kea. These observations were taken over 20.5 nights and consisted of a total of 600 spectra, of which 320 were suitable for use in tomographic reconstructions. We describe the survey design in Section \ref{sec:sd} and the data collected in Section \ref{sec:obs}. Our tomographic reconstruct techniques are described in Section \ref{sec:tomo}, and results explored in Section \ref{sec:results}.

In this paper, we assume a concordance flat $\Lambda$CDM cosmology, with $\Omega_M=0.31$, 
$\Omega_\Lambda=0.69$ and $H_0 = 70\,\kms\,\mathrm{Mpc}^{-1}$. 

\section{Survey Design and Target Selection}
\label{sec:sd}
The CLAMATO Survey covers the COSMOS field \citep{scoville:2007}, an extragalactic survey area which has excellent coverage from 
multi-wavelength and spectroscopic observational efforts, while also having a large ($\sim 1$ deg$^2$) contiguous footprint well-suited for
studying large-scale structure.
The new (post-2017) data presented within this paper follows the same target selection and observational details as the previous data release \citep{2018CLAMATO}. In this section we provide an abbreviated description of our design with a focus on any differences from the
previous iteration, and refer the reader to the earlier paper.

\begin{figure*}[ht]\centering
\includegraphics[width=0.77\textwidth,clip=true, trim=20 110 20 85]{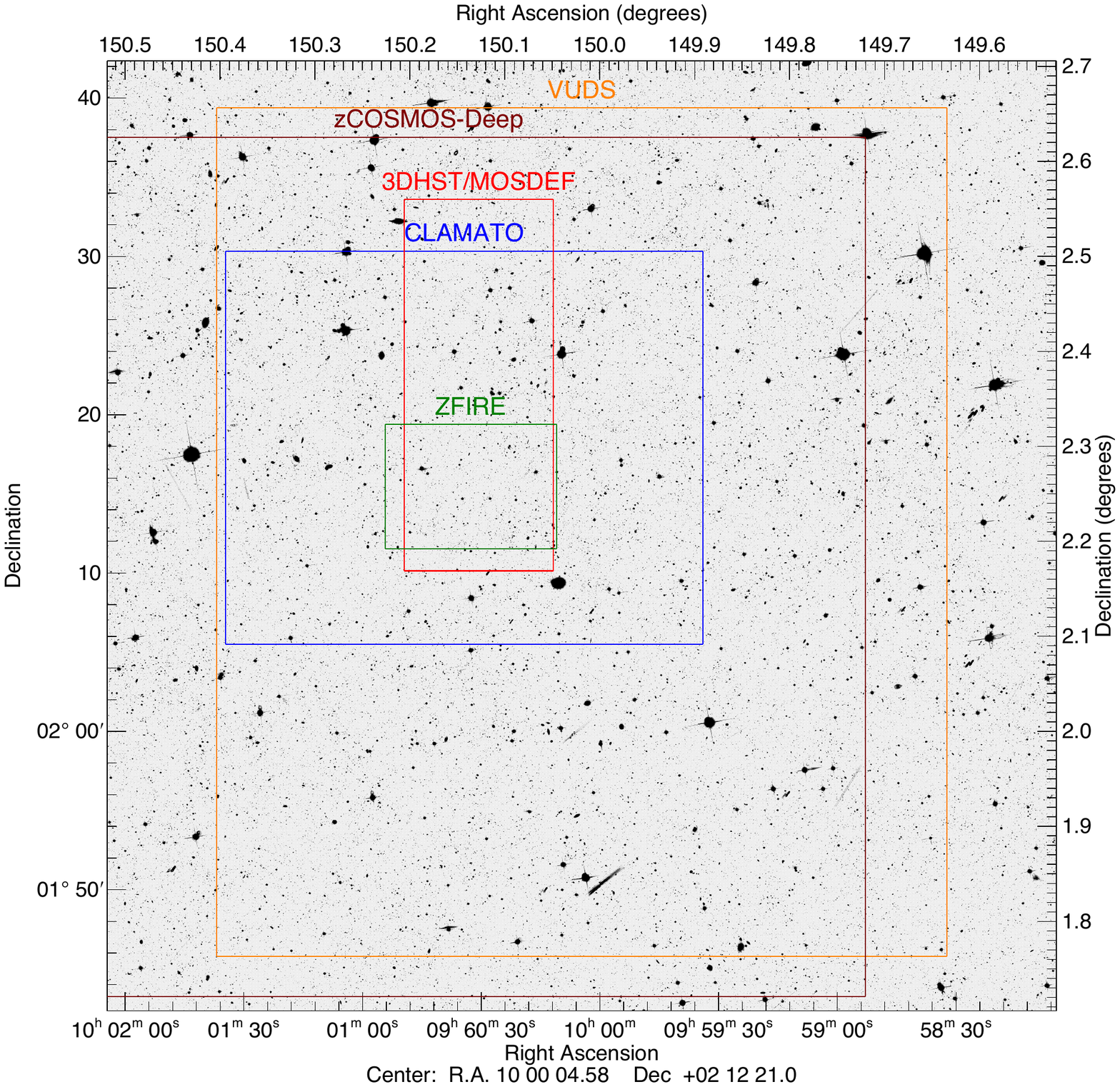}
\caption{\label{fig:survey_fields}
CLAMATO in context: the background image is the \textit{Hubble Space Telescope} ACS F814W mosaic \citep{koekemoer:2007} of the central regions in the COSMOS field, 
with the footprint of the CLAMATO \lya{} absorption tomographic map indicated in blue. Also shown are the approximate footprints for other
spectroscopic redshift surveys that probe similar redshifts, such as 3D-HST \citep{momcheva:2016} and 
MOSDEF \citep{kriek:2015} in red, zCOSMOS-Deep \citep{lilly:2007} in brown, VUDS \citep{le-fevre:2015} in orange, and ZFIRE \citep{nanayakkara:2016} in green. 
} 
\end{figure*}

 \begin{figure*}[ht]\centering
\includegraphics[width=0.72\textwidth,clip=true, trim=20 10 30 30]{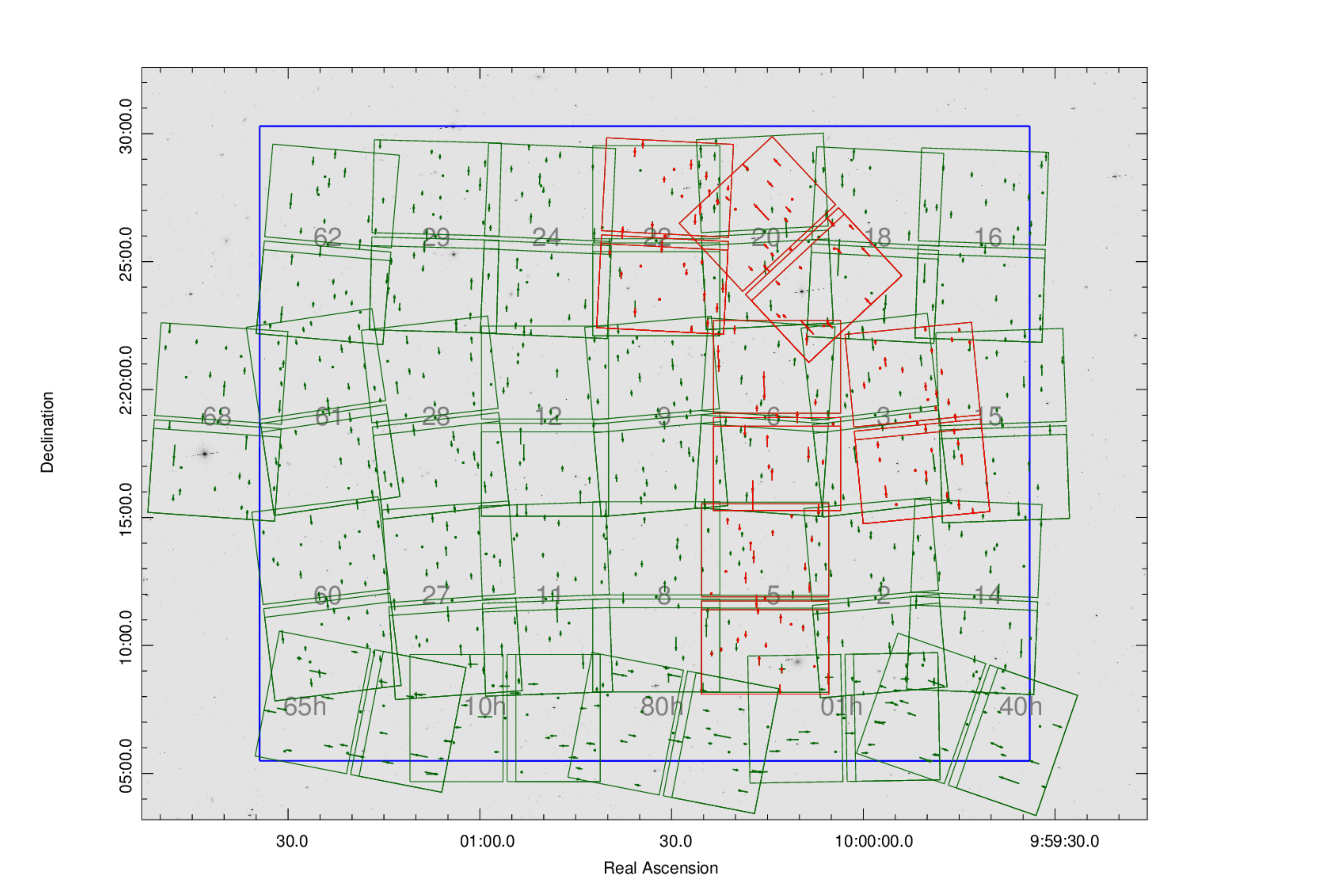}
\caption{\label{fig:slitmasks} Slits and footprints of the 32 Keck-I/LRIS slitmasks observed in CLAMATO, overlaid on top of the deep \textit{Hubble Space Telescope} ACS F814W mosaic of the COSMOS field \citep{koekemoer:2007}. The blue box indicates the footprint of the reconstructed tomographic map from the $2.15<z<2.55$ \lyaf absorption. Most of the slitmasks were designed to achieve a uniform survey layer (dark green), while several were `special' slitmasks (red) designed to obtain additional  lines of sight in specific regions. The numbers in grey are the field numbers we assigned to the slitmasks.
}
\end{figure*}

\subsection{Source Catalogs}

\label{subsec:SC}
The CLAMATO survey is designed to maximize the areal density and homogeneity of background sources in the COSMOS field \citep{scoville:2007} that could probe the \lya{} forest absorption at $z \sim 2-2.5$. To achieve this, we exploit the high-quality multi-wavelength photometric redshifts \citep{ilbert:2009, laigle:2016, davidzon:2017} available in the field, and also retarget spectroscopically-confirmed objects in the zCOSMOS-Deep \citep{lilly:2007} and VUDS \citep{le-fevre:2015} surveys\footnote{The  spectra from these surveys have coarse spectral resolution $R\sim 200$, insufficient for our purposes.}.

As a first step, we assembled a raw master catalog containing all objects $g<25.2$ within a redshift range that allows observations of the \lya{} forest in the target volume, corresponding to objects with $2.0 < z < 3.5$. We began with a compilation of spectroscopic sources in the COSMOS field (Salvato et al, in prep), which, at our redshifts of interests, are primarily from the zCOSMOS-Deep \citep{lilly:2007} and
VUDS \citep{le-fevre:2015} catalogs. This was then further supplemented with the MOSDEF spectroscopic survey \citep{kriek:2015},
 ZFIRE spectroscopic survey \citep{nanayakkara:2016}, and 3D-HST grism redshift catalog \citep{momcheva:2016}. 
 
 We then further filled in our master catalog using objects with only photometric redshifts. This initially used the $I$-band selected photometric redshift catalog from \citet{ilbert:2009}, but prior to the 2017 observing season this was supplemented 
 by near-IR selected photometric redshifts 
 \citep{laigle:2016,davidzon:2017} that yield more accurate redshift estimates.

\subsection{Selection Algorithm}

The target selection algorithm was carried out in two steps; initial prioritization of possible targets, followed by design of the individual slit-masks. Due to the spatial constraints of each LRIS slit-mask, only a subset of the possible targets could be manifested as slits.

For target prioritization, we fed the photometric/spectroscopic catalogs described in Section \ref{subsec:SC} into an algorithm designed to homogeneously probe our target area at $z\sim 2.3$. In order to ensure sufficient spatial resolution within a finite volume at $z\sim 2.3$ along the line of sight, we run our algorithm at two separate redshifts -- $\za = 2.25$ and $2.45$ -- and then merged the target lists. Our algorithm divides the survey footprint into square cells of our desired sight-line separation, 2.75 arc-minutes on each side. In each cell, we select candidate background sources that could theoretically probe the forest at \za\ in the restframe $1040\,\ang<\lambda<1216\,\ang$, i.e. between the \lya{} and Ly$\beta$ transitions, corresponding to redshifts $(1+\za) 1216/1195 -1 < \zbg < (1+\za) 1216/1040-1$. We then give the highest priority to ``bright'' sources with $g<[24.2,24.4]$ at $\za =[2.25, 2.45]$ which have confirmed spectroscopic redshifts, and down-prioritize objects with only photometric redshifts or at fainter magnitudes. 

 Due to slit-packing constraints, the algorithm deprioritizes relatively bright sources 
if another, brighter, high-confidence target is within the same cell, while fainter or photometric redshift targets might receive 
relatively high priority in the absence of other suitable background sources within its 2.75 arc-minute cell.
Due to the possibility that slit collisions from targets in other cells  
might clobber the highest-priority source within a given cell, the algorithm selected multiple sources per cell 
(with decreasing priority) where available.
This procedure selected targets as faint as $g=25.3$ in regions lacking brighter candidates, 
but such faint targets were assigned a commensurately low priority.

The initial selection of sources, and their priority rankings from this algorithm, were then fed into the AUTOSLIT3 
software\footnote{\url{https://www2.keck.hawaii.edu/inst/lris/autoslit_WMKO.html}} in order to design LRIS slitmasks.
The initial slit assignment was automatically carried out by AUTOSLIT3 based on the priorities assigned by the initial 
target selection algorithm, which we then manually refined to maximize the homogeneity of bright sources 
and the uniformity of redshift coverage
within our desired $2 \lesssim \za \lesssim 2.5$ redshift range. 
In each $7' \times 5'$ LRIS slitmask, we assigned $\sim 20-25$ science slits. Due to slit-packing constraints
and the necessity of having at least 4 alignment stars within each slitmask, 
this limited us to only $\sim 80\%$ of the high-priority targets within our desired redshift range.

We designed a uniform set of slitmasks to cover our entire survey footprint (Figure~\ref{fig:slitmasks}), but also supplemented these 
with additional slitmasks --- designed and observed in subsequent observing seasons after the initial pass--- 
to increase sightline sampling
in particular regions of interest (specifically the previously identified overdensities at $z \sim 2.5$, see section \ref{sec:results}), or to make up for shortfalls in sightline density after the initial round of observations.

Over the 2019-2020 observing season, when it became apparent that further time allocations might not be forthcoming, 
we designed a row of slitmasks with horizontal orientations to try to ensure that the overall survey footprint would end up in a roughly 
rectangular shape. These horizontal slitmasks make up the bottom row of the survey footprint as seen in Figure~\ref{fig:slitmasks}.


\section{Observations \& Data Reduction}
 \label{sec:obs}
The CLAMATO observations were carried out using the LRIS spectrograph \citep{oke:1995,steidel:2004} on the Keck-I
telescope at Maunakea, Hawai'i. The observations described in this papers were carried out in  
2014-2020 via a total time allocation of 20.5 nights, of which 18.5 nights were allocated by the University of California Time Allocation Committee (TAC) and 2 nights
were from the Keck/Subaru exchange time given by the National Astronomical Observatory of Japan TAC.
Out of this overall allocation, we achieved approximately 85hrs of on-sky integration.

CLAMATO focused on the LRIS blue channel, covering the range of $3700\,\ang < \lambda < 4400\,\ang$ which corresponds to the restframe \lya{} at $2.1\lesssim \za \lesssim2.6$. We used the 600-line grism blazed at $4000\,\ang$ in the blue channel which, with the $1''$ slitwidth we used, has a spectral resolution of $R \equiv \lambda/\Delta\lambda \approx 1100$. This corresponds to a FWHM of $\approx 4\,\ang$ which translates to a line-of-sight spatial resolution of $3\,\hMpc$ at $z\sim2.3$, comparable to the separation between our  lines of sight. We use the red channel to assist in redshift determination and object identification.

Observations were conducted with a mean seeing of FWHM~$\approx 0.7''$, and in $<0.8''$ seeing we typically exposed for a total of 7200s per typical survey slitmask. In suboptimal seeing we increased exposures up to a total of 14400s in order to attain approximately homogenous minimal signal-to-noise over the sky. For survey slitmasks designed to plug gaps, we integrated up to 19800s to gain sufficient signal-to-noise on fainter background sources and to deal with poor seeing conditions during those observations. Individual exposures in the blue channel were typically 1800s, while those in the red channel were only 860s to mitigate
for the effects of cosmic ray hits on the red-channel's thick fully-depleted CCDs \citep{rockosi:2010}.

The gathered data was reduced using the LowRedux routines from the XIDL software 
package\footnote{\url{http://www.ucolick.org/~xavier/LowRedux}}. We performed initial flat-fielding, slit definition, and sky subtraction, and then co-added the 2D images of the individual exposures. We then traced the 1D spectra, additionally co-adding 1D spectra from different observing epochs which could not be co-added in 2D.

\begin{figure}
\includegraphics[width=0.5\textwidth]{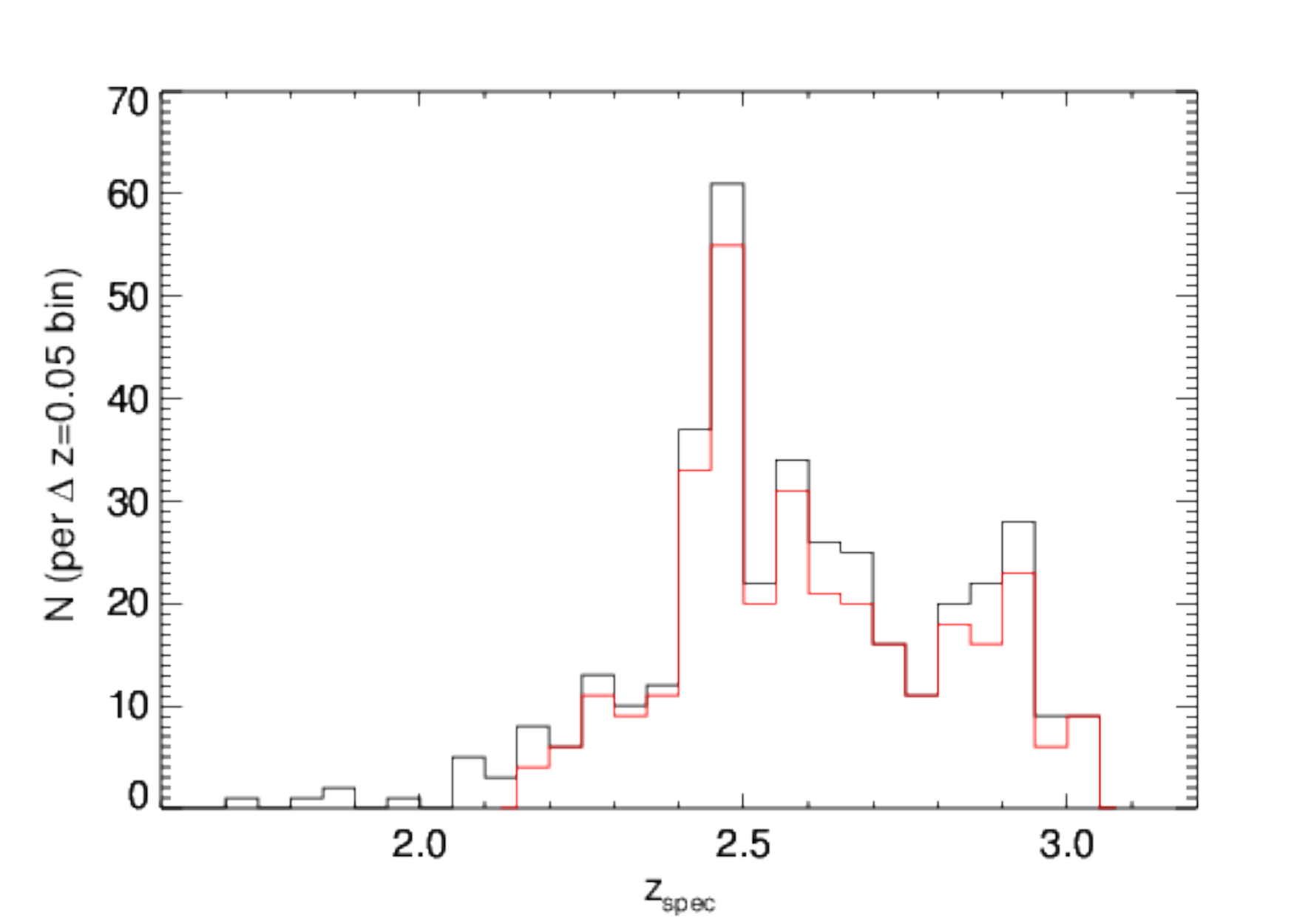} 
\caption{\label{fig:source_zhist}
Redshift distribution of well-identified ($\geq 3$ confidence rating) spectra from CLAMATO, shown
as the black histogram with redshift bins of $\Delta(z)=0.05$. 
The red histogram indicades background sources that were actually used to tomographicaly reconstruct
 the foreground \lyaf{} at $2.05<\za<2.55$.  These plot axes leave out 11 objects at $z<1.6$ and 1 object at $z>3.2$. Note that the spike at $z\sim2.45$ is due to previously identified large over-density in this area \citep{diener:2015, chiang:2015,casey:2015,wang:2016,2018A&A...619A..49C}.
}
\end{figure}

\begin{figure}\centering
\includegraphics[width=0.48\textwidth]{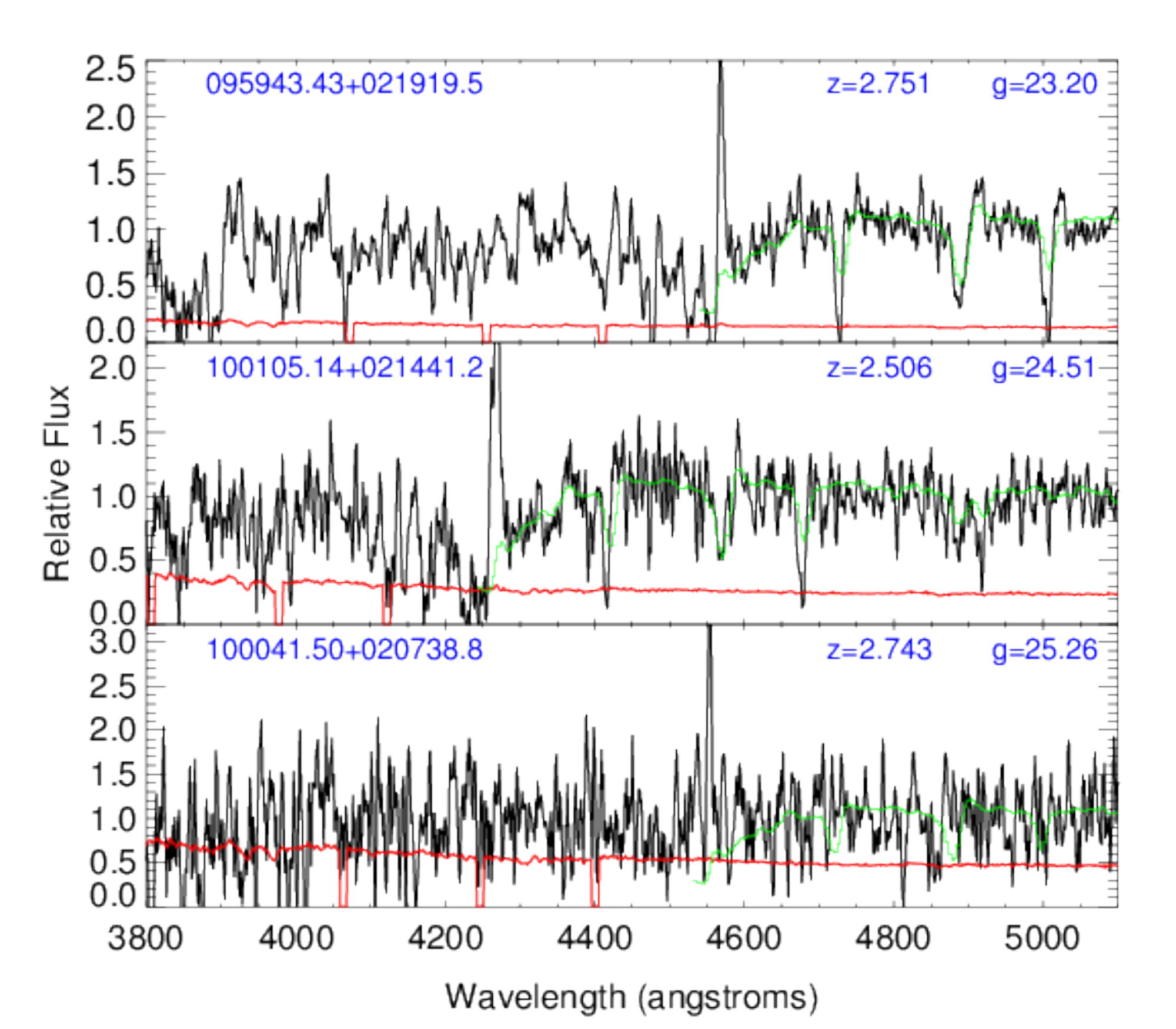}
\caption{\label{fig:spec_eg}
Examples of the reduced spectra used for IGM tomography in CLAMATO, showing, from top to bottom, galaxies with 
the brightest, median, and faintest $g$-magnitude. All three objects, coincidentally, exhibit clear \lya{} emission. For clarity, the spectra have been smoothed with a 3-pixel tophat filter.
The noise spectrum is shown in red, while redwards of \lya{} we overplot the \citet{shapley:2003} LBG spectral template.
} 
\end{figure}

From the 32 unique slitmasks observed in the 2014-2020 CLAMATO campaign, we reduced and extracted
600 spectra from the blue channel, not including 19 spectra from unrelated `filler' programs. The resulting spectra were then visually inspected and compared with common line transitions and spectral templates (particularly the \citet{shapley:2003} composite LBG template) to determine their redshift and classification type. Each spectra is then assigned a confidence flag with ranking of 0-4, where 0 denotes no attempt at identification due to a corrupted spectra or little source flux, 1 is a rough best guess, 2 is a low-confidence redshift, 3 is an object with reasonable enough classification certainty to be used for scientific results, while 4 is a high confidence redshift determined from multiple spectral features. Out of all the spectra, 393 spectra were determined to have $\geq$3 confidence rating of which 377 were at redshifts of $z>2$ (Figure~\ref{fig:source_zhist}). The faintest object of this category was a $g=25.29$ galaxy at $z=2.568$, although this was a \lya{} emitter that could not be used for IGM tomography. 
 These high-redshift, high confidence sources were classified into 359 galaxies ($95.5\%$) and 18 broad-line quasars (4.5\%). Both populations are suitable for \lyaf{} absorption surveys, but require different continuum fitting methods, as described in \citealt{lee:2018}. We show a full table of extracted sources in Table~\ref{tab:tomo_obj}, while examples of the spectra are shown in Figure~\ref{fig:spec_eg}. Nearly all sources with identified continuum were used for our end tomographic analysis, see \S~\ref{sec:tomo}.



 

 \begin{deluxetable*}{l c c c c c c c c c c c c}[htb!]
\tablecolumns{13}
\tablecaption{\label{tab:tomo_obj} CLAMATO Data Release 2 Redshift Catalog}
\tablehead{
ID\# & $\alpha$ (J2000)\tablenotemark{a} & $\delta$ (J2000)\tablenotemark{a} & $g$-mag\tablenotemark{a} & 
$z_{\rm photo}$\tablenotemark{b} & $z_{\rm spec}$ 
& Conf\tablenotemark{c} & Type & $t_{\rm exp}$ (s) & Tomo\tablenotemark{d} & $\mathrm{S/N_{Ly\alpha1}}$\tablenotemark{e} &
$\mathrm{S/N_{Ly\alpha2}}$\tablenotemark{f}  & $\mathrm{S/N_{Ly\alpha3}}$\tablenotemark{g} }
\startdata
00762 &  10 01 00.905 &  +02 17 27.96 & 24.21 &   1.11 &  2.465 & 2 & GAL & 
7200
 & N & \nodata & \nodata & \nodata \\
00765 &  10 01 00.297 &  +02 17 02.58 & 24.64 &   2.93 &  2.958 & 4 & GAL & 
7200
 & Y & \nodata & \nodata &  2.5 \\
00767 &  10 01 14.934 &  +02 16 45.23 & 24.73 &   0.21 &  2.578 & 3 & GAL & 
12600
 & Y &  3.1 &  3.2 &  3.0 \\
00771 &  10 01 06.870 &  +02 16 23.38 & 24.70 &   2.58 &  2.530 & 3 & GAL & 
7200
 & Y &  1.6 &  1.9 &  2.0 \\
00780 &  10 01 14.359 &  +02 15 15.84 & 24.28 &   0.08 &  0.082 & 2 & GAL & 
7200
 & N & \nodata & \nodata & \nodata \\
00783 &  10 01 07.412 &  +02 14 58.31 & 24.27 &   2.59 &  2.579 & 4 & GAL & 
10200
 & Y &  4.1 &  4.5 &  4.7 \\
00784 &  10 01 15.952 &  +02 14 48.41 & 22.02 &   2.47 &  2.494 & 4 & QSO & 
9000
 & Y & 11.5 & 13.1 & 22.1 \\
00785 &  10 01 05.138 &  +02 14 41.21 & 24.51 &   2.44 &  2.506 & 4 & GAL & 
10200
 & Y &  2.1 &  2.5 &  2.8 \\
00787 &  10 01 21.083 &  +02 14 16.48 & 24.41 &   2.62 &  2.491 & 3 & GAL & 
9000
 & N &  0.8 &  1.0 &  1.1 \\
00788 &  10 01 33.860 &  +02 14 25.19 & 24.24 &   2.62 &  2.738 & 3 & GAL & 
9000
 & Y & \nodata &  1.6 &  1.8 \\
\enddata
 \tablenotetext{a}{Source positions and magnitudes from \citet{capak:2007}.}
 \tablenotetext{b}{Photometric redshift estimate; see text for details.}
 \tablenotetext{c}{Redshift confidence grade, similar to that described in \citet{lilly:2007} but without fractional grades.}
 \tablenotetext{d}{Usage in \lya{} forest tomographic reconstruction}
 \tablenotetext{e}{Median per-pixel spectral continuum-to-noise ratio within the $2.05<\za<2.15$ \lyaf{}.}
  \tablenotetext{f}{Median per-pixel spectral continuum-to-noise ratio within the $2.15<\za<2.35$ \lyaf{}.}
   \tablenotetext{g}{Median per-pixel spectral continuum-to-noise ratio within the $2.35<\za<2.55$ \lyaf{}.}
 \tablecomments{Table 1 is published in its entirety in the machine-readable format.
      A portion is shown here for guidance regarding its form and content.}
\end{deluxetable*}

\section{Tomographic Reconstructions}\label{sec:tomo}

\begin{figure*}[ht]\centering 
\includegraphics[width=0.72\textwidth,clip=true, trim=0 10 30 30]{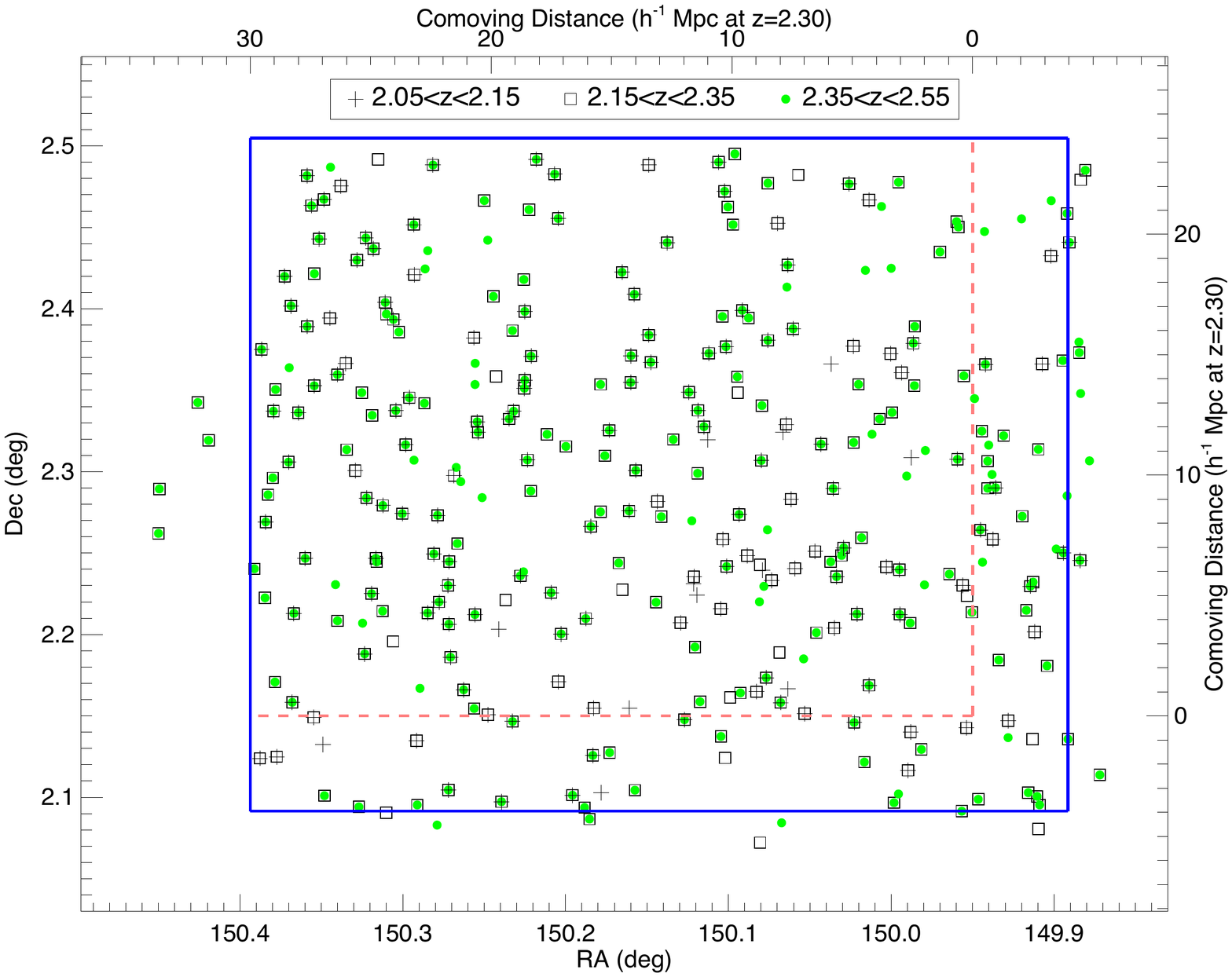}
\caption{\label{fig:sightlines_pos}
Points indicate the celestial positions of the \lyaf{}  lines of sight used to tomographically reconstruct the \lyaf{} at $2.05<z<2.55$, 
with the blue rectangle showing the angular footprint adopted for the tomographic reconstruction --- the pink dashed 
rectangle is the footprint from our previous data release \citep{lee:2018}. 
The different symbols denote coverage over different redshift ranges.
Some background sources have the correct redshift to cover large ranges of our targeted foreground
redshift range
and are therefore indicated by multiple symbols.  The top and right-hand axes
denote the coordinates of our tomographic map grid in units of transverse comoving distances, evaluated
at the mean redshift of our tomographic map..
} 
\end{figure*}

\begin{figure}
\includegraphics[width=0.48\textwidth,clip=true,trim=10 0 0 0]{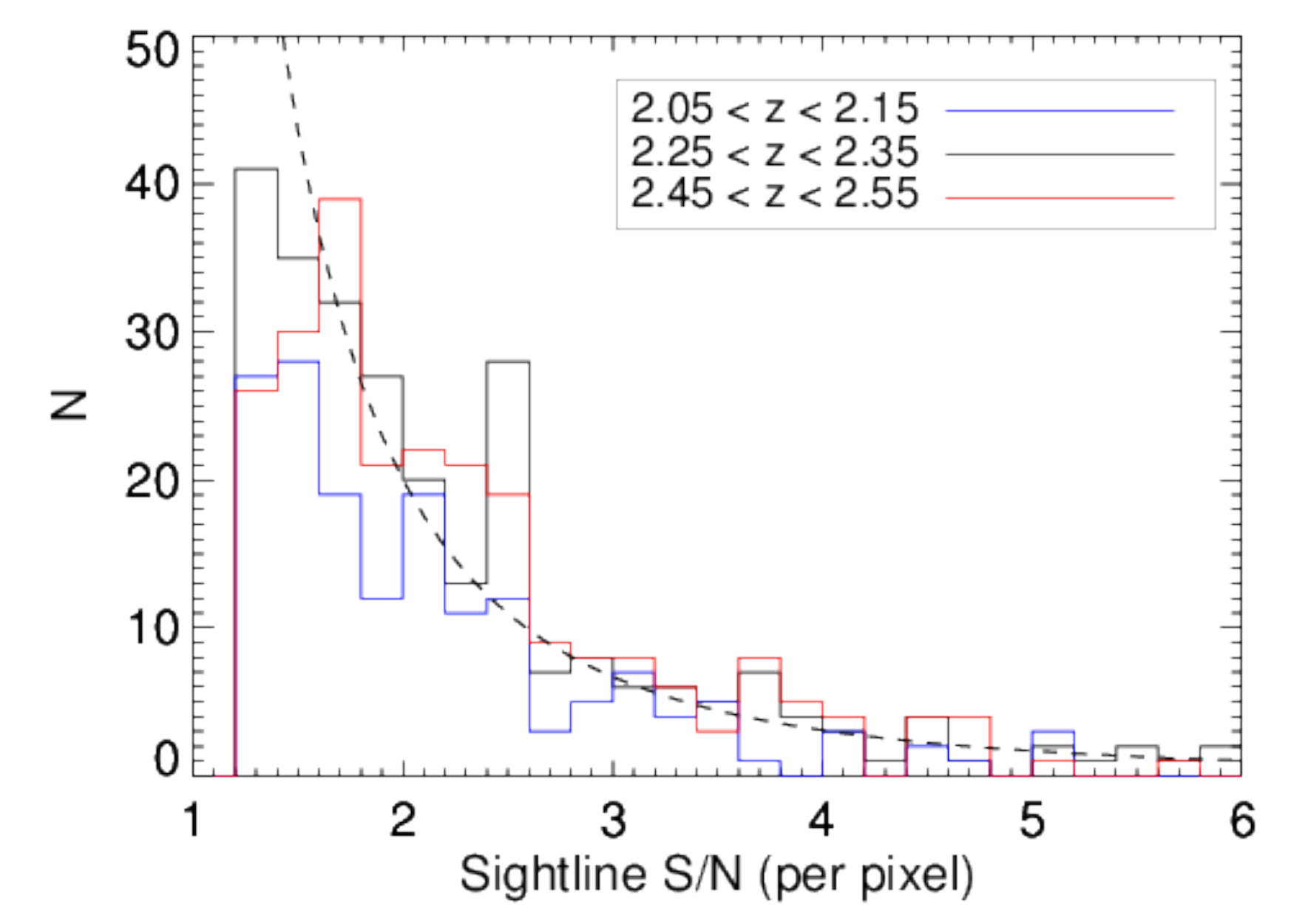}
\caption{\label{fig:snhist}
Distribution of the median sightline signal-to-noise within the \lyaf{}, evaluated at several redshift bins 
within our map volume. A small number of higher signal-to-noise  lines of sight have been left out by these plot axes. 
The dashed curve is a power-law with index of $-2.7$, which is a reasonable approximation for our signal-to-noise
distribution.
}
\end{figure}

In this section, we describe our methods for 3D tomographic reconstruction of the \lya{} absorption seen in the spectra of our
background selection. First, we describe the pre-processing of the reduced 1D spectra and setup of the tomographic reconstruction
grid. Then, we describe the two methods we use for tomographic reconstruction: (i) Wiener-filtering of the transmitted \lya{} flux, which was also used in our previous
papers (e.g., \citealt{lee:2014a, lee:2016, lee:2018}), and (ii) a constrained realization method to estimate the underlying matter density field \citep{horowitz:2019,horowitz:2020}.

\subsection{Data Preparation}

We selected reduced spectra based on their continuum-to-noise ratio (CNR), defined as the value of the fitted continuum divided by the noise, within the \lya{} forest absorption wavelengths for our tomographic analysis. We evaluate this ratio over three distinct absoprtion redshift windows:  $2.05<\za<2.15$, $2.15<\za<2.35$ and $2.35 < \za < 2.55$. All high redshift objects with estimated confidence $\geq 3$ and $\langle \mathrm{CNR}\rangle \geq 1.2$ over any of these absorption windows were used in our tomographic reconstructions. This approach was fairly aggressive, selecting the vast majority of objects which had a confident redshift estimate (Figure~\ref{fig:source_zhist}), only excluding objects which had negligible continua. We show the sky positions of the selected lines of sight  in Figure~\ref{fig:sightlines_pos}.

 \begin{figure}
\includegraphics[width=0.5\textwidth]{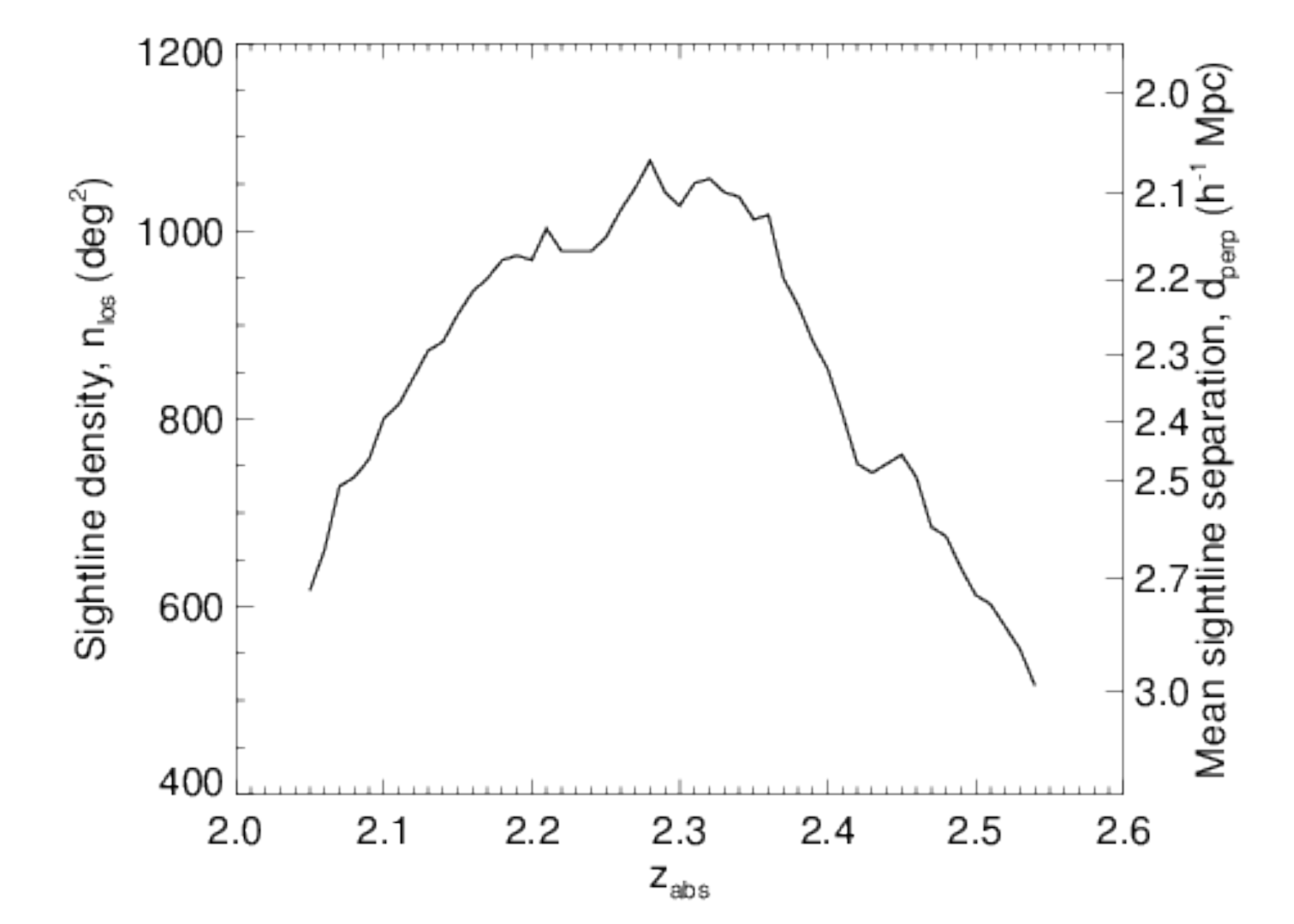} 
\caption{\label{fig:dperp_zdist}
Effective area density of Lyman-$\alpha$ forest  lines of sight over the redshift range of the CLAMATO tomographic
reconstruction. The right axis labels the equivalent mean separation between  lines of sight, $\dperp$, evaluated at $z=2.3$.
The peak sightline density is $1056\,\deg^{-2}$ at $\za = 2.28$, corresponding to $\dperp = 2.07\,\hMpc$. Note that this estimate does not take into account the pixel masks in the spectra.}

\end{figure}

A total of 320 spectra from our observations fulfilled both the redshift criteria and signal-to-noise classification to contribute to our tomographic reconstruction of the \lyaf{} within the target redshift range $2.05<\za<2.55$ in our footprint. We show the distribution of the estimated \lyaf{} signal-to-noise in Figure~\ref{fig:snhist} at different redshift ranges within our volume. We can model our signal-to-noise distribution with a power-law with index of $-2.7$, as done in \citet{krolewski:2017,krolewski:2018}. Based on the sky positions of the  lines of sight, the transverse footprint of the tomographic reconstruction spans a comoving region of $30.1' \times 24.8'$  in the R.A.\ and declination dimensions, respectively (see Figure~\ref{fig:sightlines_pos}), centered at $10^h00^m34\fs21, +02\degr17'53.49''$ (J2000). This angular size is equivalent to a comoving scale of $30\,\hMpc \times 24\,\hMpc$ at $\langle z \rangle=2.30$. The mean sighline density is $\langle n_\mathrm{los}  \rangle = 856 \,\mathrm{deg}^{-2}$ when averaged over the entire volume, equivalent to $\dperp = 2.35\,\hMpc$ at $z=2.3$. At the two redshift extremes, $z=[2.05,2.55]$, we find an effective sightline density of $n_\mathrm{los} = [617, 515]\,\mathrm{deg^{-2}}$, corresponding to an average transverse comoving seperation of $\dperp = [2.73, 2.98] \,\hMpc$ (see Figure~\ref{fig:dperp_zdist}). Towards the middle of our redshift range we have the highest effective sight-line density, peaking at $1075\,\mathrm{deg^{-2}}$ at  $\za = 2.28$, equivalent to a transverse comoving separation of $\dperp = 2.07\,\hMpc$. The overall sightline densities in the current CLAMATO data is marginally worse than in DR1, as the boost from the known galaxy overdensities at $z\sim 2.5$ in DR1 has been diluted by the increased map area. 



The \lyaf{} fluctuation at each pixel can be calculated from a given spectra with an estimated continuum, $C$. We divide the observed spectral flux density, $f$, by the  mean \lyaf{} transmitted flux, $\langle F \rangle (z)$, multiplied by the estimated continuum value;
\begin{equation}\label{eq:delta_f}
\delta_F = \frac{f}{C\;\langle F \rangle (z) } - 1.
\end{equation}
We adopt the \citet{faucher-giguere:2008a} values for $\langle F \rangle (z)$, which are based on a sample of 86 high-resolution, high-signal-to-noise quasar spectra with absorption features extending well over this redshift range.

The intrinsic continua, $C$, of a quasar source is estimated based on a PCA-based mean-flux regulation 
\citep[MF-PCA; e.g.,][]{lee:2012a, lee:2013}. For each spectrum, a continuum template is fitted to determine the correct shape for the QSO emission lines. A linear function is further fitted within the \lyaf{} region (restframe $1041\,\ang<\lambda<1185\,\ang$ ) to ensure that the mean absorption is consistent with \citet{faucher-giguere:2008a}. 
An analogous procedure is applied to the galaxy spectra, although assuming a broader \lyaf{} range ($1040\,\ang<\lambda<1195\,\ang$) and fixing the continuum template from \citet{berry:2012}. For galaxies we also explicitly remove intrinsic metal absorption lines at rest-frame  \waveion{N}{2}{1084}, \waveion{N}{1}{1134}, \waveion{C}{3}{1176},
and \ion{Si}{2} $\lambda\lambda 1190, 1193$ using a $\pm7.5\,\ang$ (observed frame) mask. Uncertainties in the continuum estimation are propagated assuming a $\sim 10\%$ rms for the noisiest spectra ($\mathrm{S/N} \sim 2$ per pixel) and $\sim 4\%$ rms for the highest signal spectra ($\mathrm{S/N} \sim 10$ spectra); see \citet{lee:2012a} for additional discussion.

We use the resulting flux pixel values, $\delta_F$, and associated noise uncertainties, $\sigma_N$, as inputs for our tomographic reconstructions. These extraced values are made publicly available and described in greater detail in Appendix~\ref{app:dr}.

To perform our tomographic reconstructions, we define a cartesian grid in our footprint (Figure~\ref{fig:sightlines_pos}). To simplify our analysis, we assume a fixed Hubble parameter, $H(z)$, throughout our volume evaluated at  $\langle z \rangle = 2.30$. This is equivalent to setting a fixed differential comoving distance $\mathrm{d}\chi/\mathrm{d}z$, allowing for a direct correspondence between redshift segment length $\delta z$ and comoving distance $\delta\chi$. Our angular footprint of $30.1' \times 24.8'$ translates to a fixed comoving scale of $34\,\hMpc \times 28\,\hMpc$ at all redshifts in our
 map. Note that this will result in a slight variation of our smoothing kernel's physical size from one side of our redshift range to the other. The alternative to this procedure would be use an evolving $H(z)$ over the volume, resulting in outward flared skewer positions relative to the comoving grid; this was found to negligible effect on the cosmic void analysis of \citet{krolewski:2017} while also breaking the one-to-one correspondence between $[x,y]$ and [RA, Dec], complicating analysis of the maps. 
 
 The angular footprint of this grid is $\sim 30 \%$ larger than \cite{2018CLAMATO} and $4.5\times$ larger than that in \citet{lee:2016}. 


\subsection{Wiener Filtering}

To construct maps using the \lyaf{} flux alone, we use a Wiener filtering (WF) scheme as done in \cite{2018CLAMATO}. The basic algorithm is described in \citet{pichon:2001} and \citet{caucci:2008}, and we use an implementation developed in \citet{stark:2015a}\footnote{\url{https://github.com/caseywstark/dachshund}}. Other methods for direct flux reconstruction also exist including \citealt{cisewski:2014} and \citealt{2021ApJ...916...20L}, which could be applicable depending on the end use of the maps. For WF, we solve for the \lyaf{} flux fields:
\begin{equation}\label{eq:wiener}
\delta^{\mathrm{rec}}_F=\cmd\cdot (\cdd+\mathbf{N})^{-1}\cdot\delta_F,
\end{equation}
where $\cdd+\mathbf{N}$ and $\cmd$ are the data-data (with noise) and map-data covariances, respectively. We use a preconditioned conjugate gradient method to approximately solve the combination of matrix inversion and multiplication. For the noise covariance, we assume a diagonal form of $\mathbf{N}\equiv N_{ii}=\sigma_{N,i}^2$ which ignores sub-dominant errors from continuum mis-identification and other correlated errors. We impose a noise-floor of $\sigma_{N,i}\geq 0.2$ in order to not overly weight a small subset of high signal-to-noise lines of sight.

We also assume a Gaussian data-data covariance of the form
$\cdd=\cmd=\mathbf{C(r_1,r_2)}$, following \cite{caucci:2008}, and 
\begin{equation}  \label{eq:kernel}
\mathbf{C(r_1,r_2)}=\sigma_F^2\exp\left[-\frac{(\Delta r_\parallel)^2}{2L^2_\parallel}\right]\exp\left[-\frac{(\Delta r_\perp)^2}{2L^2_\perp}\right],
\end{equation}
where $\Delta r_\parallel$ and $\Delta r_\perp$ are the distance between 
$\mathbf{r_1}$ and $\mathbf{r_2}$ along, and transverse to the line-of-sight, respectively. We adopt a transverse and line-of-sight
correlation lengths of $L_\perp= 2.5\,\hMpc$ and $L_\parallel=2.0\,\hMpc$, respectively, 
as well as a normalization of $\sigma^2_F=0.05$. This formulation of the covariance matrix and parameters were found in \citet{stark:2015a} to reasonably approximate the true covariance, have no explicit cosmological dependence, and provide accurate reconstructions on mock CLAMATO catalogs. The choices of $L_\perp$ and $L_\parallel$ are set by the average sight-line separation and the LOS spectral smoothing, respectively.

From the input map data of 84608 pixels, we perform a Wiener filter reconstruction using the \citet{stark:2015a} algorithm which solves the matrix inversion using a pre-conditioned conjugation gradient solver with a stopping tolerance of $10^{-3}$. The run-time for the entire volume was approximately 
1000s using a single core of a 
Apple MacBook Pro laptop with 2.9 GHz Intel Core i5 processors and 16GB of RAM. 

Our output grid is of size $68 \times 56 \times 876$ cells each $0.5\,\hMpc$ on a side with an effective smoothing scale of $\sim 2-3\,\hMpc$. The resulting map is publicly available for download as a binary file; 
see Appendix~\ref{app:dr} for details.

\subsection{Density Reconstructions and Cosmic Structure}
\label{subsec:tardis}
In addition to the Wiener Filter flux reconstructions, we also reconstruct the underlying matter density field using the Tomographic Absorption Reconstruction and Density Inference Scheme (TARDIS, \citet{horowitz:2019}). This method iteratively solves for the initial density field which gives rise to the observed structures through a forward modelling framework. 
Unlike the Wiener filtered reconstructions, this method solves for the underlying dark matter density fields, instead of the \lya{} flux maps. 
This assumes an underlying $\Lambda$CDM cosmology and an analytical approximation to map from the evolved dark matter density to \lyaf{}  optical depth. One can view this process as iteratively solving for an intial density field, $\delta_m^i$, given the following forward operating steps;
\begin{equation}
    \delta_m^i \xrightarrow[\textrm{N-Body }]{} (\delta_m^{z_\alpha},v_m^{z_\alpha}) \xrightarrow[\textrm{FGPA}]{} (\tau_\textrm{real},v_m^{z_\alpha}) \xrightarrow[\textrm{RSD}]{} \tau_\textrm{red}
\end{equation}
where from the end redshift space optical depth, $\tau_\textrm{red}$, we select matching lines of sight as the CLAMATO Survey and iterate to find the density field that
best matches the CLAMATO data. We assume the fluctuating Gunn-Peterson approximation, $\tau = A \exp{\delta_m^\gamma}$, for the hydrodynamical mapping from the matter density to \lya{} flux, which was found to cause minimal decrease in cosmic structure recovery performance \citep{horowitz:2020}. For our N-Body solver, we use a particle mesh scheme as described in \cite{2020arXiv201011847M}. This resulting $\tau_\textrm{red}$ field is compared along each line of sight with the true data with a Gaussian likelihood weighted by the inferred pixel noise. All steps of our forward operator are differentiable, allowing fast optimization using first order based methods. In this case we use a Limited Memory Broyden–Fletcher–Goldfarb–Shanno (L-BFGS) solver which approximates the Hessian information via a truncated recursion relation from the previous update's gradients. 

In \citet{horowitz:2020}, this framework was further extended to allow joint reconstruction of \lyaf{} tomography with an overlapping galaxy sample. 
This work found a synergistic reconstruction effect, where \lyaf{} absorption traced diffuse structures on large scales while galaxies were able to reconstruct the amplitudes of over-densities. 
As CLAMATO is located in the COSMOS field, there are a number of possible galaxy surveys to use for a joint reconstruction including ZFIRE, zCOSMOS, MOSDEF, etc. However, large scale structure reconstruction requires a well defined radial selection function which is difficult to determine in cases where the survey is designed to study a particular structure or galaxies of a certain sub-population (see \cite{ata:2020} for a discussion). We therefore use zCOSMOS \citep{lilly:2007} as the corresponding galaxy sample to combine with the \lyaf{} absorption, which comprises $\sim 45\%$ of all galaxies in the overlapping volume and has a well characterized selection function which we take from \citet{ata:2020}.

In practice it would be computationally difficult to reconstruct the full volume at once due to the elongated geometry. We divide the volume into four overlapping cubic subvolumes of $L=128\,\hMpc$ each, with 15 \hMpc\, overlap in
the radial dimension. In addition we offset the data 10 \hMpc\, into each reconstructed subvolume in order to minimize possible edge effects due to the simulations periodic boundary conditions. The reconstructed subvolumes are then pasted together, with a linear weighted averaging in the overlapping region. We have verified with mock catalogs that this pasting scheme does not introduce artifacts into the simulated volume nor decrease reconstruction quality in the overlap. Our reconstruction with this method takes approximately 30 minutes on a NVIDIA Tesla V100 PCle GPU card for the entire survey volume. We provide publicly our resulting reconstructed maps in numpy file format gridded in 1 \hMpc\ cells.

As the reconstruction recovers directly the dark matter field in both real and redshift space, we can use these recovered fields for direct classification of the cosmic web environment. We use the pseudo-deformation tensor as described in \citet{lee:2016a,krolewski:2017} and based on work in \citet{hahn:2007,bond:1996,forero-romero:2009}.
The pseudo-deformation tensor is defined as the Hessian of the gravitational potential, capturing the curvature information of the smoothed underlying density field. 
This tensor can be efficiently computed with known matter density in Fourier space, $\delta_k$, as
\begin{equation}
    \tilde{D}_{ij} = \frac{k_i k_j}{k^2}\delta_k.
    \label{eq:diften_k}
\end{equation}
The eigenvectors of this tensor are related to the principle inflow  or outflow directions depending on if the corresponding eigenvalue is positive or negative. The signature of the eigenvalues themselves can be used to classify the resulting cosmic structures as nodes, filaments, sheets, and voids (see Table \ref{tab:signature}).

\begin{table}[]
    \centering
    \begin{tabular}{c|c|c|c}
         & $\lambda_1$ & $\lambda_2$ &$\lambda_3$\\
         \hline
         Node & +  & +  & + \\
         Filament & +  & +  & - \\
        Sheet & +  & -  & -\\
        Void & -  & -  & -\\
    \end{tabular}
    \caption{Signature of the eigenvalues of the deformation tensor for structure classification.}
    \label{tab:signature}
\end{table}

\section{Results}\label{sec:results}

\begin{figure*}\centering
\includegraphics[width=\textwidth, clip=true, trim=0 0 0 0]{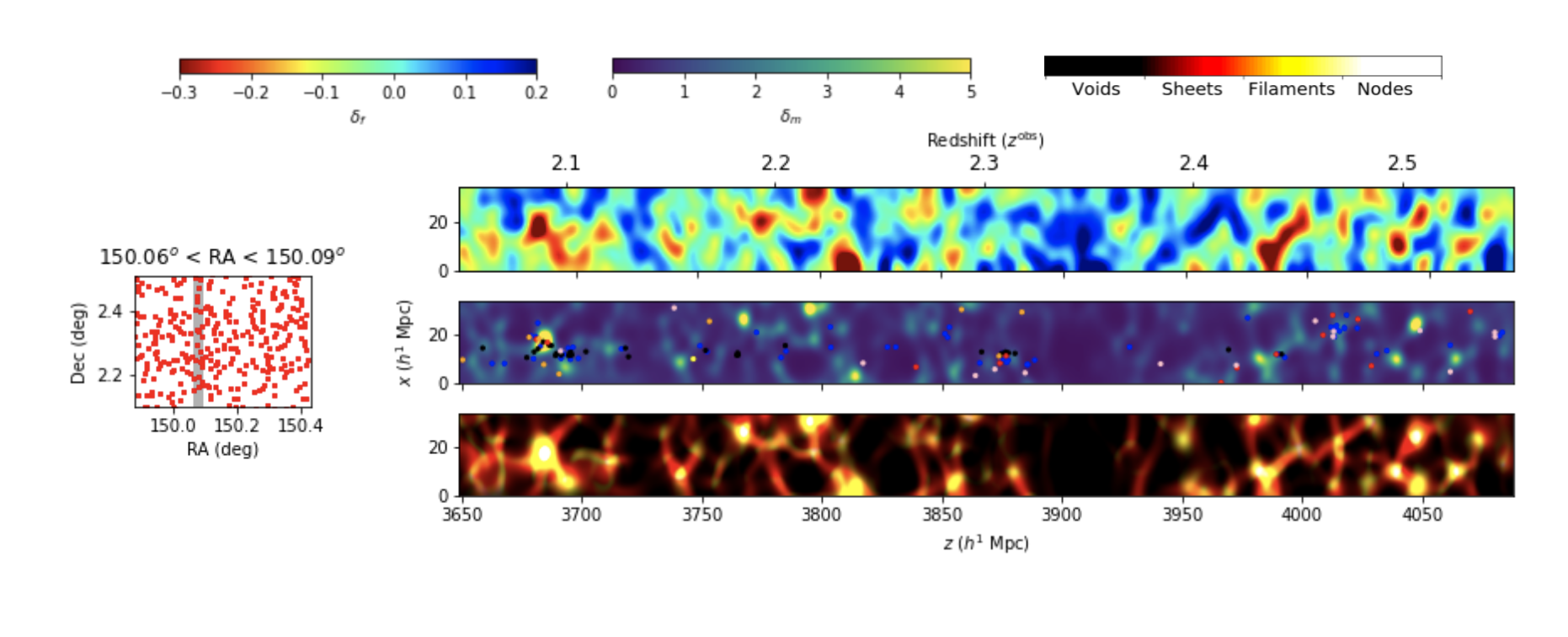}
\caption{\label{fig:slicemaps}
Wiener-filtered tomographic reconstructions of the \lyaf{} absorption field, $\delta_F^{\mathrm{rec}}$,
 at $2.05<\za<2.55$ from the
current CLAMATO data (color map), shown after smoothing with an isotropic $R=2\,\hMpc$ Gaussian kernel. Below the Wiener-filtered reconstruction, we show the inferred dark matter distribution from the TARDIS reconstruction and the inferred cosmic web classification, shown as a continuous color spectrum for the eigenvalues which can be mapped onto a discrete classification.
Each color panel shows the field projected over a $2\,\hMpc$ R.A. slice, 
the position of which is denoted by the shaded region in the subpanels to the left that also show the sightline
positions on the sky as red dots. The color convention for the absorption is such that red denotes overdensities 
while blue denotes underdensities. In the middle panel we show the overlapping galaxies within the slice; blue dots from MOSDEF, 
black dots from ZFIRE, pink dots from VUDS, orange dots from zCOSMOS-Deep, and red dots from CLAMATO. 
This sequence is continued in Figure \ref{fig:a_slicemaps}, \ref{fig:b_slicemaps}, and \ref{fig:c_slicemaps}
}
\end{figure*}

\begin{figure*}[th]\centering
\includegraphics[width=0.65\textwidth,clip=true, trim=0 0 0 0]{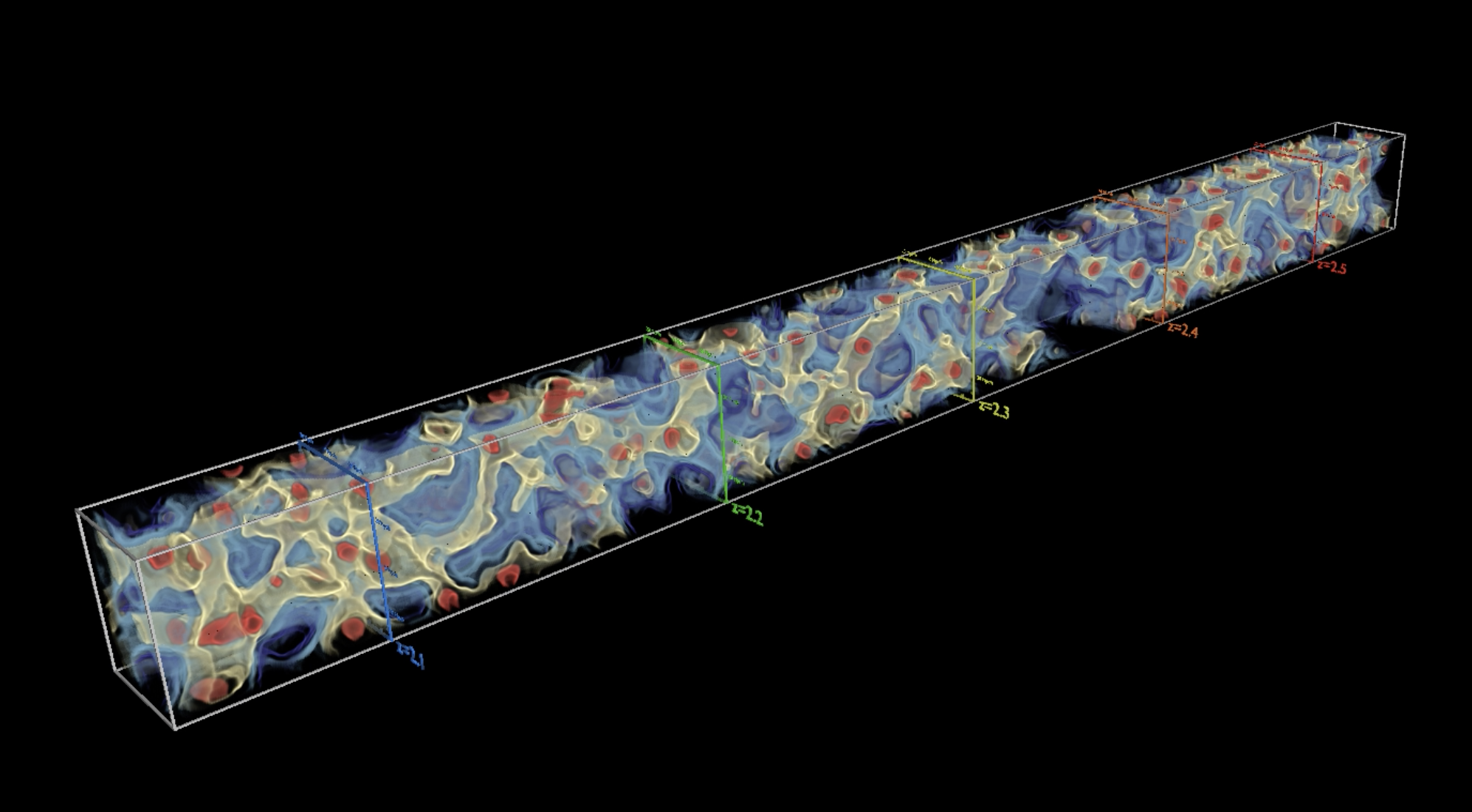}
\caption{\label{fig:screenshot}
Still image from our 3D video visualization of the CLAMATO reconstructed cosmic web
(smoothed with a $R=2\,\hMpc$ Gaussian kernel). Each cosmic structure classification is indicated with a different color; red for nodes, yellow for filaments, blue for sheets, and transparent for voids.  These videos are available in the online journal and online at \url{https://bit.ly/clamatoDR2}.
}
\end{figure*}

\begin{figure*}\centering
\includegraphics[width=0.65\textwidth,clip=true, trim=0 0 0 0]{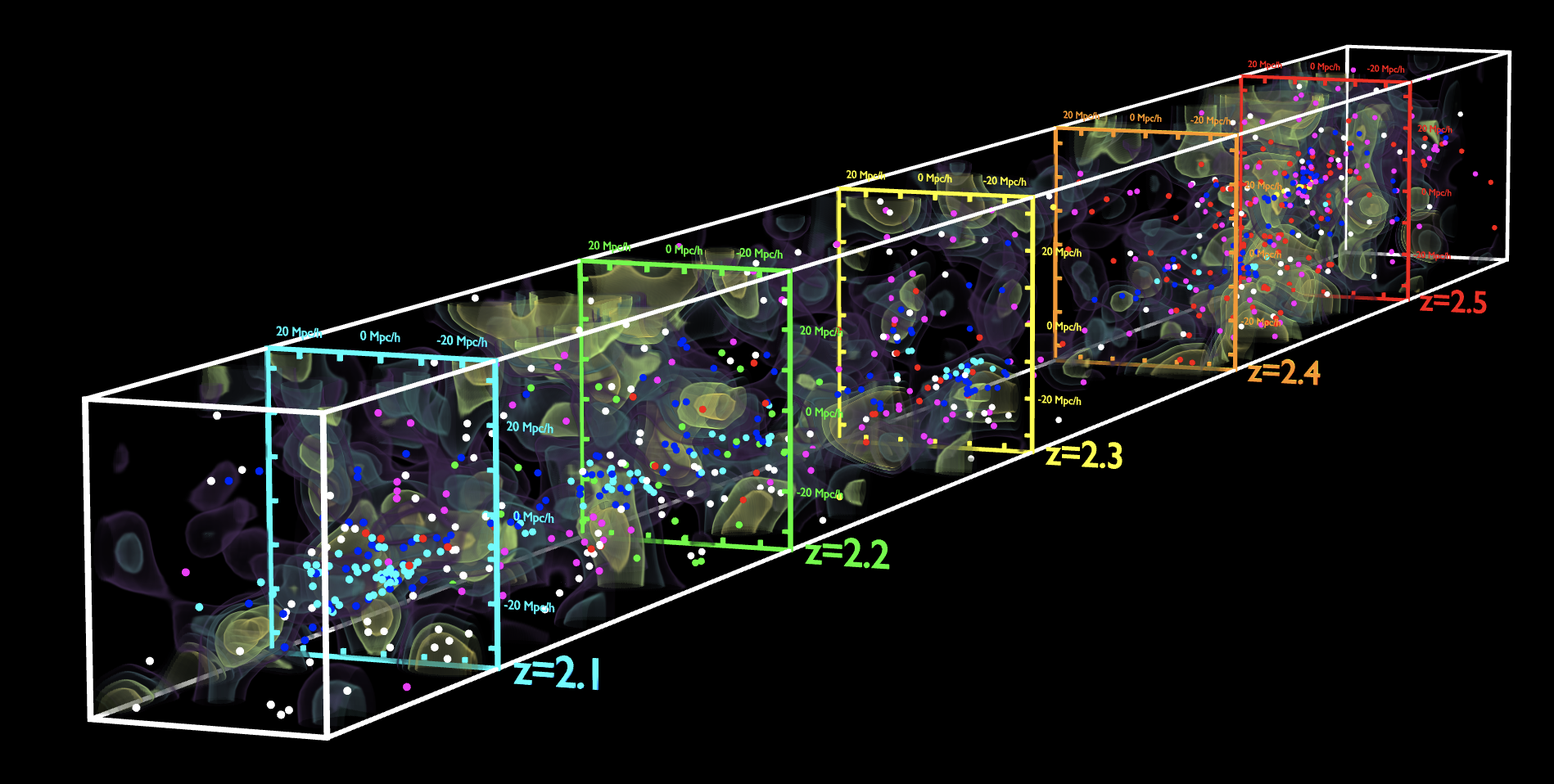}
\caption{\label{fig:x3d}
Three-dimensional rendering of the CLAMATO tomographic map, showing isodensity contours at $\deltarec = -0.260$, $-0.20$,$-0.14$,$-0.09$ along with coeval galaxy positions shown as dots color coded by galaxy survey (zCosmos: white, VUDS: purple, MOSDEF: blue, zfire: teal, CLAMATO: red). This figure is available online as an
interactive figure (\url{https://www2.mpia-hd.mpg.de/homes/tmueller/projects/clamato2020/clamato_vol.html}). 
By left-clicking and moving the mouse, 
the viewpoint can be rotated, while the right-mouse button or scroll wheel can be used to zoom in or out; 
double left-clicking at any point in the map focuses the viewpoint there. A number of options are available to toggle different galaxy surveys and change the display of the flux iso-contours. 
}
\end{figure*}

We show an example slice of our reconstructed map via both Wiener Filtering and via forward model reconstruction in Figure~\ref{fig:slicemaps}. In addition to the reconstructed maps, we show a visualization of the cosmic structure as defined by the eigenvalues of the reconstructed matter deformation tensor, as well as coeval spectroscopically observed galaxies in the volume from different surveys. Each slab is projected over a thickness of $2\,\hMpc$, along  right ascension, with the $y$-axis of each plot representing change in declination and the $z$ axis representing the redshift or line-of-sight dimension. For a clear comparison, we smooth all maps with a Gaussian kernel of $R=2\,\hMpc$, the approximate average sight-line spacing of the survey. We have also created a number of three dimensional renderings of our reconstructed flux field, dark matter density field, and inferred cosmic structures. These are available both in movie form  (Figure~\ref{fig:screenshot}) as well as a manipulable interactive  figures (Figure~\ref{fig:x3d}).

There are a number of potential applications of these reconstructed maps to constrain intergalactic medium gas properties, galaxy formation and evolution, and cosmology. However, in this section we focus on a qualitative discussion of the structures recovered and will explore other applications in future work.

 \subsection{Large-Scale Structure Features}
 
 All of the visualizations show a rich cosmic web environment which is apparent in both the Wiener Filtered maps as well as the forward modelled reconstructions. We see a close correspondence between the reconstructed flux maps and the underlying galaxy surveys, with the significant caveat that all overlapping surveys have a non-uniform selection function along the line of sight and on the sky. Note that due to the varying selection functions and sky coverage of each survey, this galaxy sample is incomplete over our volume and it would be challenging to infer cosmic structures from galaxies alone \citep{horowitz:2020}.
 
 Compared with \citet{2018CLAMATO}, the expanded footprint allows the reconstruction of a complete picture of the cosmic web. With a transverse length of 35 \hMpc\ and 30 \hMpc, we are able to completely capture filamentary structures spanning identified cosmic nodes within our survey area, as well as resolve sheet structures between filaments (see Fig. \ref{fig:screenshot}). In addition to filamentary structures, as in \citet{2018CLAMATO}, we see apparent voids and (proto)clusters which have been previously discussed in past data releases in the same field (see \citealt{lee:2015} and \citealt{krolewski:2017}) but now with increased sky coverage. We see that DR2 completely includes a number of previously identified galaxy protoclusters including the $z\sim2.1$ galaxy protocluster first identified through the ZFOURGE medium-band 
photometric redshift survey \citep{spitler:2012}
and later confirmed with NIR spectroscopy \citep{nanayakkara:2016}. 

In addition, the previously identified  $z\approx 2.5$ overdensity comprised of the $z=2.44$ protocluster \citep{diener:2015, chiang:2015}, $z=2.47$ protocluster \citep{casey:2015}, and
X-ray detected $z=2.51$ cluster  \citep{wang:2016,2018A&A...619A..49C}, all appear to form a massive interconnected structure (see Figure~\ref{fig:screenshot} and \ref{fig:x3d}) extending from $2.44 < z < 2.52$. While this structure was truncated by the survey geometry in \citet{2018CLAMATO}, it does not appear to extend significantly beyond our current survey volume.  The VUDS galaxy survey has been used to infer the general shape of this structure (see \citealt{2018A&A...619A..49C}), and we find our inferred Wiener filtered flux recovers a similar qualitative shape (see second panel of Figure \ref{fig:b_slicemaps}). The exact correlation on small scales ($\sim 1$ h$^{-1}$ Mpc) between galaxy density and \lyaf{} flux depends on the galaxy feedback model \citep{sorini:2018}. The late time fate of these mega-structures is of significant interest, with \citet{wang:2016} arguing that the $z=2.51$ overdensity, in itself, 
might collapse into a $2\times 10^{15}\,\mathrm{M}_\odot$. Using forward modelling techniques \citep{horowitz:2019,horowitz:2020,ata:2020} we will explore the late time fate of these structures in future works (Ata et al. in Prep).
\subsection{Comparison of TARDIS and Wiener Filter Reconstruction}

TARDIS is a fundamentally different approach that Wiener filter reconstruction since the resulting density fields are constrained to evolve from gravitational evolution and the hydrodynamical properties are determined via the Fluctuating Gunn Peterson Approximation, see Section \ref{subsec:tardis}. Substantial differences in the field reconstructed between these two methods could be indication of strong deviations from simple analytical approximations for the IGM flux-density relationship due to strong feedback within clusters environments (see, e.g., \cite{2022arXiv220110169K}), or, e.g., possible galaxy quenching in overdense environments \citep{newman:2022}. In the case of mock catalogs which include hydrodynamical effects, TARDIS has shown to be still highly accurate in classification of cosmic structures at the resolution available to the CLAMATO survey \citep{horowitz:2020}. Also see \citet{horowitz:2019} for a discussion of TARDIS reconstruction vs. Wiener Filtering in the absence of hydrodynamical effects.

\begin{figure}
    \centering
    \includegraphics[width=0.48\textwidth]{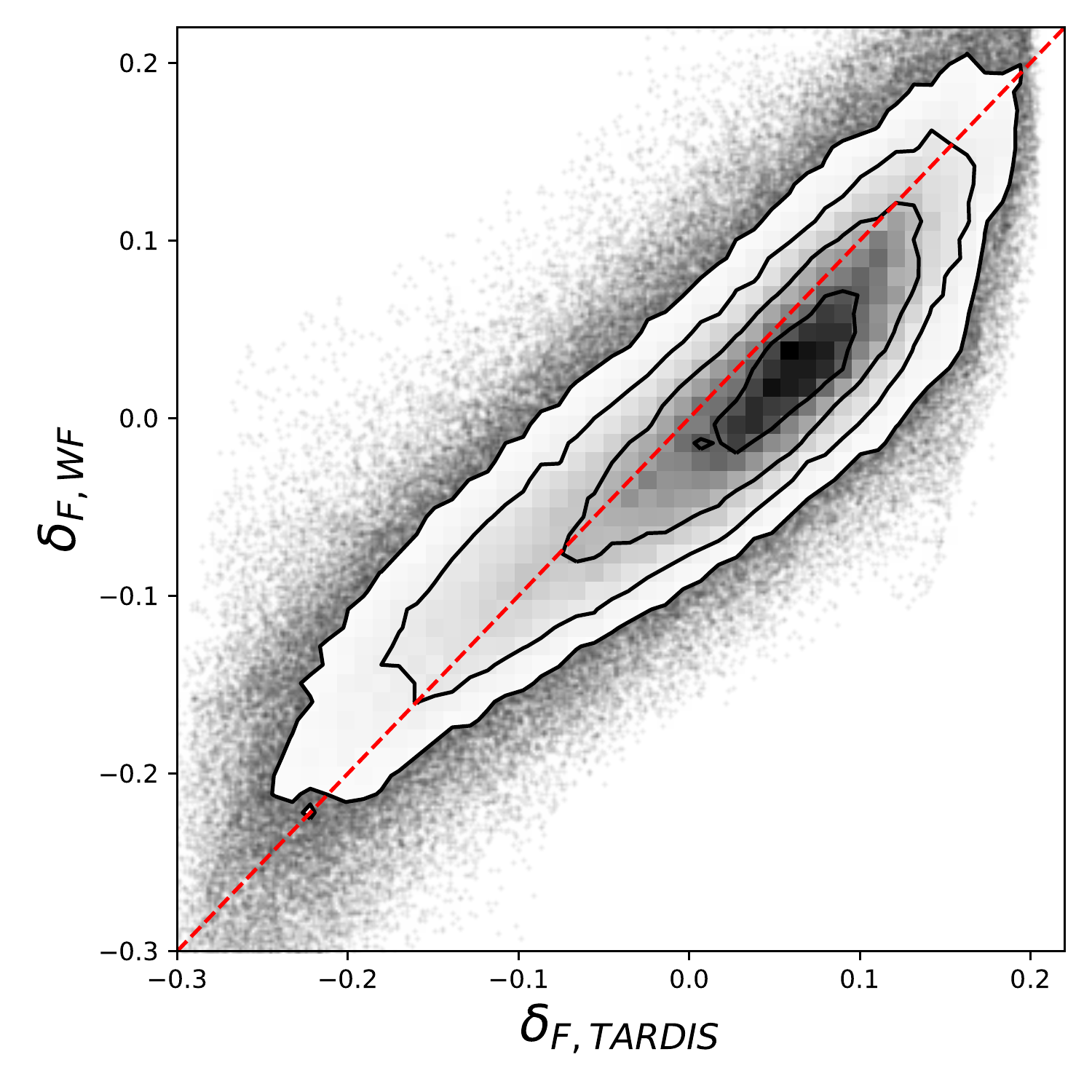}
    \caption{Relationship between the inferred flux from the TARDIS reconstruction and from the Wiener Filter reconstruction, both smoothed by 2 \hMpc. We find no significant biases between the two maps.}
    \label{fig:TARDIS_WF}
\end{figure}

In addition to the slice-maps shown in Figures \ref{fig:a_slicemaps},\ref{fig:b_slicemaps} \ref{fig:c_slicemaps}, we show the relationship between the inferred flux from TARDIS, calculated by performing an FGPA-transformation on the resulting density field, and that inferred from the Wiener filter approach in Figure \ref{fig:TARDIS_WF}. We find the results are largely consistent, with Pearson Correlation coefficient of 0.86. Further propagating this analysis to cosmic structure as defined by the eigenvalues of the deformation tensor, we find a Pearson Correlation coefficient of 0.83 between structures classified using the Wiener Filter and TARDIS, indicating similar cosmic structure reconstruction. Further investigation into the causes of the differences between these maps will be the focus of future work.

\subsection{Three-dimensional Visualizations}

Figure \ref{fig:screenshot} shows a 3d visualization of the CLAMATO reconstructed cosmic web rendered using a self-developed software based on the Open Graphics Library (OpenGL) and the OpenGL Shading Language (GLSL)\footnote{OpenGL 4.6 Specification, \url{https://www.khronos.org/opengl}} and implemented in C++. 

Because our tomographic map consists only of scalar values, we can apply direct volume rendering where each density value is mapped to a particular color and opacity value via a transfer function. Then, a ray is generated for each image pixel and is sampled at regular intervals throughout the volume where the mapped RGBA values from the transfer function are composited in front-to-back order. For a smoother representation we use tricubic interpolation.

To highlight the cosmic structure classification, we use four Gaussians of nearly equal width to generate the transfer function. Sampling artifacts are reduced by making use of the pre-integration technique \citep{klaus:2001}.

Finally, mesh and volume rendering are combined via two-pass rendering, where we first render color and depth of the grid mesh into two textures.In the second pass, the depth texture is used while the rays are integrated to check for intersections with the mesh.

Figure \ref{fig:x3d} shows a 3d visualization of the CLAMATO tomographic map also using direct volume rendering. 
This time the transfer function consists of four very thin Gaussians to represent isodensity contours.
The interactive online version (\url{https://www2.mpia-hd.mpg.de/homes/tmueller/projects/clamato2020/clamato_vol.html}) is based on WebGL/threejs. The isodensity values as well as the corresponding opacities can be modified via slider controls in the online interface.

\section{Conclusion}
\label{sec:conclusion}
In this paper, we have described the second data release of the CLAMATO Survey, the first systematic attempt
at implementing 3D \lyaf{} reconstruction on several-Mpc scales using high area densities ($\sim 1000\,\deg^{-2} $) of background LBG and quasar spectra. With Keck-I LRIS observations of 23 multi-object slitmasks over $\sim 0.2 \mathrm{deg}^2$ in the COSMOS field, we obtained 393 spectra with confident redshifts, of which 320 spectra were suitable for use in tomographic reconstruction in the target redshift range, $2.05< \za  <2.55$ \lyaf{}. This set of  lines of sight has an average transverse seperation of only $\dperp = 2.35\,\hMpc$, allowing us to create a three dimension tomographic map of the IGM and also infer the underlying dark matter density field. 
From this reconstructed map, we can identify for the first time the detailed cosmic web (including filaments and sheets) at high redshifts $z\sim 2.4$. We have made all catalogs, pixel data, and reconstructed maps available to the general public along with various analysis code (see Appendix~\ref{app:dr}).
With the extended volume of DR2, we are currently performing multiple additional analyses in this volume, including studying galaxy properties as a function of IGM and cosmic environment, analysis of the protoclusters within the volume, intrinsic alignments of galaxies with cosmic structure, constraints on hydrodynamical properties of the IGM from the various reconstructed maps, and cross-correlations between  \lyaf{} and coeval galaxies.

Going forward, the analysis done with the CLAMATO survey will serve as a test for planned and ongoing surveys over a wider sky region. The high redshift component of the Galaxy Evolution survey with the Prime Focus Spectrograph (PFS) on the 8.2m Subaru Telescope \citep{sugai:2015} will perform an IGM tomographic analysis with similar sightline densities and noise properties as the CLAMATO survey over a $\sim 40\times$ larger cosmic volume. Beyond PFS, there are various additional facilities in different stages of planning ( e.g. the 11.25m Maunakea Spectroscopic 
Explorer \citep[MSE,][]{mcconnachie:2016a}) which will offer multiplex factors of several thousand over $\sim 1\deg^2$ fields-of-view allowing IGM tomography over tens or hundreds of square degrees. While the cosmological constraining power of the current CLAMATO survey is limited, with the expanded footprint of PFS or MSE there are potential applications to detect the  weak-lensing of the
\lyaf{} \citep{metcalf:2017} as well as perform full three dimensional \lyaf{} power spectrum analysis \citep[e.g.,][]{2018JCAP...01..003F,2003ApJ...585...34M}.

\section*{Acknowledgements}
BH is supported by the AI Accelerator program of the Schmidt Futures Foundation.
KGL acknowledges support from JSPS Kakenhi Grants JP18H05868 and JP19K14755.  MA was supported by JSPS Kakenhi  Grant  JP21K13911.  
  We are also grateful to the entire COSMOS collaboration for their assistance and helpful discussions.
The data presented herein were obtained at the W.M. Keck Observatory, 
which is operated as a scientific partnership among the California Institute of Technology, 
the University of California and the National Aeronautics and Space Administration (NASA). 
The Observatory was made possible by the generous financial support of the W.M. Keck Foundation.
  The authors also wish to recognize and acknowledge the very significant cultural role and reverence that the summit of Maunakea has always had within the indigenous Hawai'ian community.  We are most fortunate to have the opportunity to conduct observations from this mountain.

\appendix

\section{Data Release}\label{app:dr}

The second data release of the Keck-CLAMATO data is publicly available on Zenodo.\footnote{\url{https://doi.org/10.5281/zenodo.7524313}}
These include the reduced spectra, continuum-normalized
 \lyaf{} pixels used as the input for the tomographic reconstruction, the tomographic map of the $2.05<z<2.55$ IGM, and the inferred density field and cosmic structure from TARDIS.
 
 There are a total of 600 reduced spectra in the data release, available in FITS format with the following headings:
 \begin{itemize}\setlength\itemsep{0.1pt}
 \item {HDU0:} Object spectral flux density, in units of $10^{-17} \mathrm{ergs\,s^{-1}\,cm^{-2}\,\ang^{-1}}$
\item {HDU1:} Noise standard deviation
\item {HDU2:} Pixel Wavelengths in angstroms
 \end{itemize}
 On the data release webpage, we will provide an ASCII catalog that contains the information in Table~\ref{tab:tomo_obj} 
 as well as the filenames for the spectra of each object. Note that the released spectrophotometry, especially in the red, might be unreliable.
 
We also will provide a binary file with the intermediate product of 84608 concatenated \lyaf{} pixels (Equation~\ref{eq:delta_f})
at $2.05<\za<2.55$ from the background sources that satisfy our redshift and signal-to-noise criteria.
We will release the $\delta_f$ values and associated pixel noise as a function of the sky positions in $[x,y,z]$ on our tomographic map grid. The sky coordinates, $x$ and $y$, correspond to the transverse comoving distances in the R.A. and Declination directions, respectively. We choose an origin location of
$[\alpha_0, \delta_0] = [9^h 59^m 47\fs999, +02\degr9'0.00"]$ (J2000) or $[\alpha_0, \delta_0] = [149.9500\degr, 2.1500\degr]$.  
The $z$ coordinate corresponds to line-of-sight comoving distance relative to the origin redshift of $\za=2.05$. 
As mentioned in Section~\ref{sec:tomo}, we use a fixed conversion between redshift and the comoving distance in our volume. With our choice of cosmology, evaluated at $\langle z \rangle = 2.30$, we have $\chi = 3874.867\,\hMpc$ and
 $\mathrm{d}\chi/\mathrm{d}z = 871.627\,\hMpc$. This intermediate binary file is the primary input used for the Wiener reconstruction and TARDIS algorithms to create the three dimensional tomographic maps.

\begin{figure*}\centering

\includegraphics[width=0.80\textwidth]{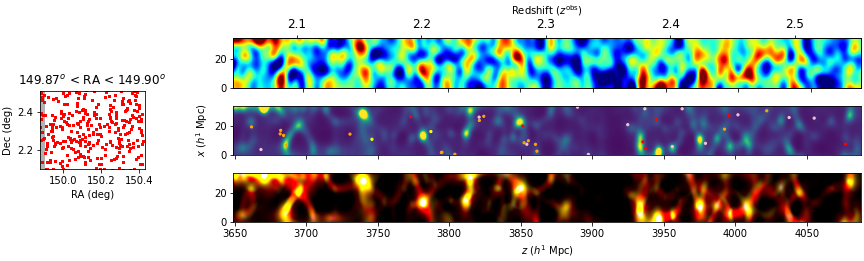}

\includegraphics[width=0.80\textwidth]{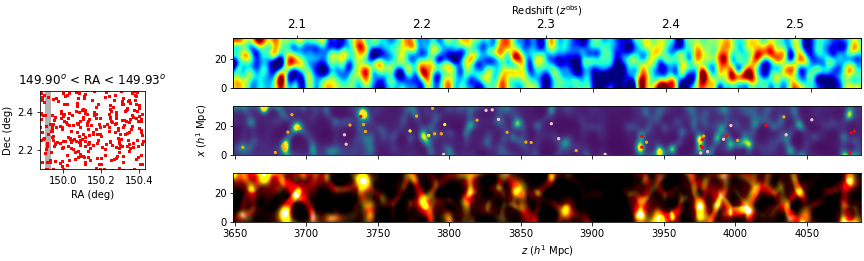}

\includegraphics[width=0.80\textwidth]{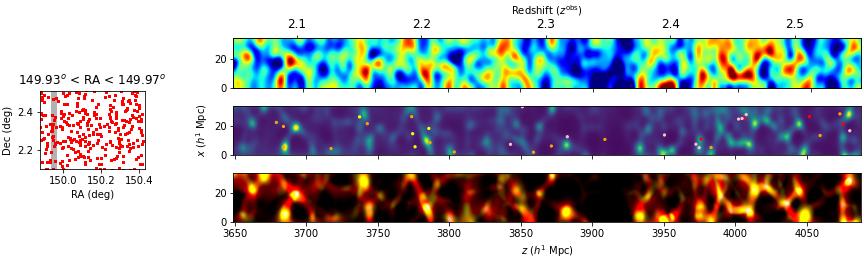}

\includegraphics[width=0.80\textwidth]{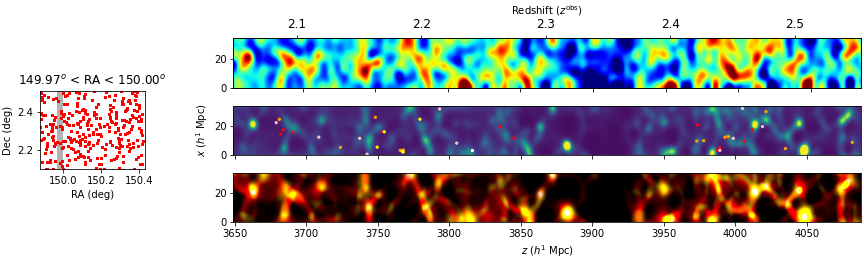}

\includegraphics[width=0.80\textwidth]{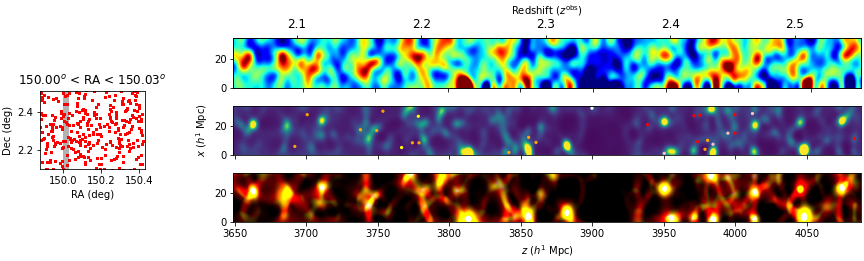}

\caption{\label{fig:a_slicemaps}
Wiener-filtered tomographic reconstructions of the \lyaf{} absorption field, $\delta_F^{\mathrm{rec}}$,
 at $2.05<\za<2.55$ from the
current CLAMATO data (color map), shown after smoothing with an isotropic $R=2\,\hMpc$ Gaussian kernel. Below the Wiener-filtered reconstruction, we show the inferred dark matter distribution from the TARDIS reconstruction and the inferred cosmic web classification, shown as a continuous color spectrum for the eigenvalues. Each color panel shows the field projected over a $2\,\hMpc$ Dec slice,
the position of which is denoted by the shaded region in the subpanels to the left that also show the sightline
positions on the sky as red dots. In the middle panel we show the overlapping galaxies within the slice; blue dots from MOSDEF, 
black dots from ZFIRE, pink dots from VUDS, orange dots from zCOSMOS-Deep, and red dots from CLAMATO. 
See Fig \ref{fig:slicemaps}, continued on Fig \ref{fig:b_slicemaps} and \ref{fig:c_slicemaps}
}
\end{figure*}

\begin{figure*}\centering

\includegraphics[width=0.85\textwidth]{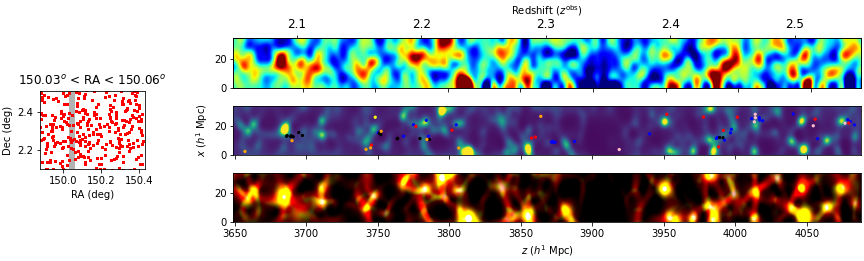}

\includegraphics[width=0.85\textwidth]{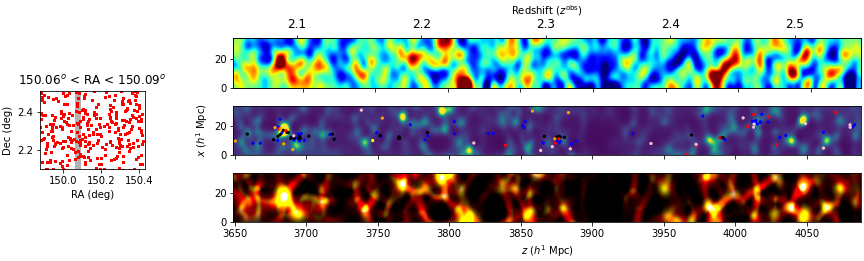}

\includegraphics[width=0.85\textwidth]{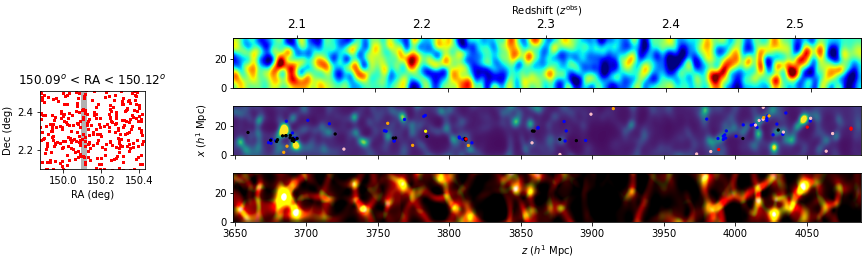}

\includegraphics[width=0.85\textwidth]{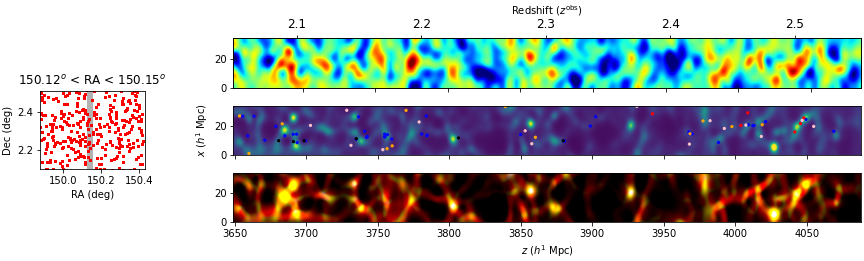}

\includegraphics[width=0.85\textwidth]{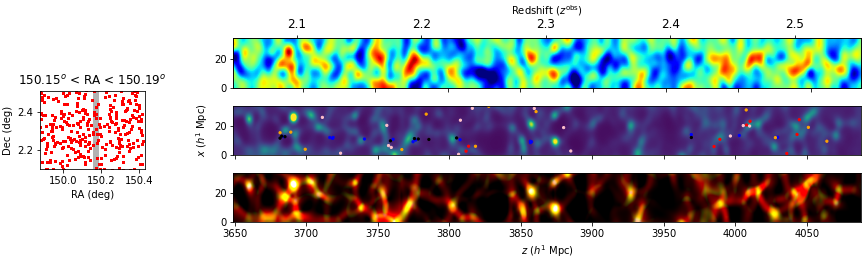}

\caption{\label{fig:b_slicemaps}
Continued from Figure \ref{fig:a_slicemaps}.
}
\end{figure*}

\begin{figure*}\centering

\includegraphics[width=0.85\textwidth]{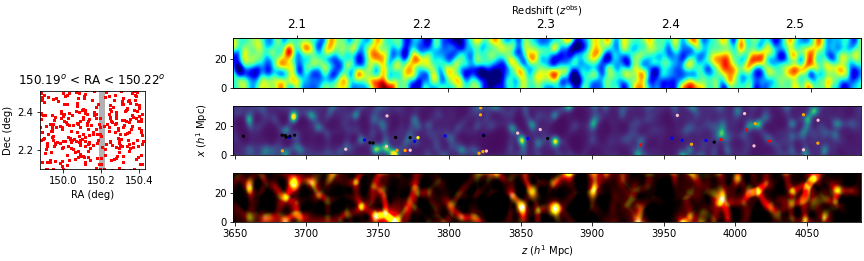}

\includegraphics[width=0.85\textwidth]{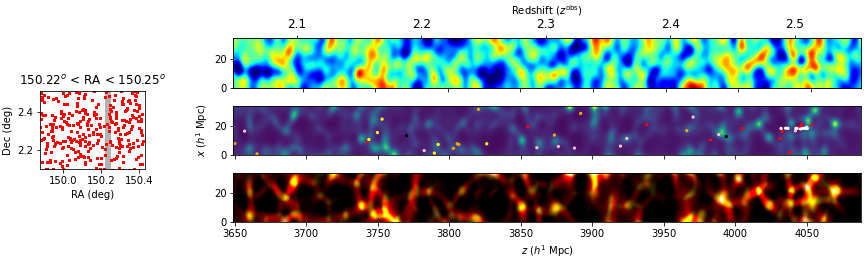}

\includegraphics[width=0.85\textwidth]{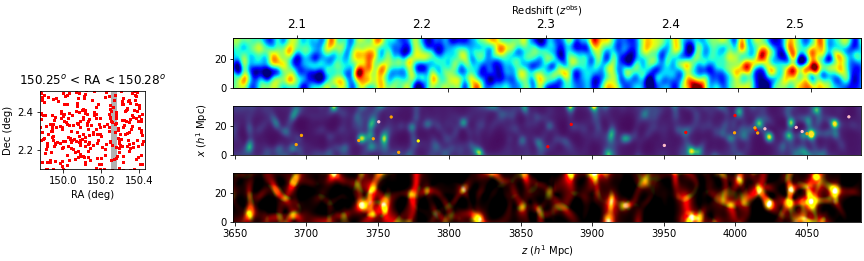}

\includegraphics[width=0.85\textwidth]{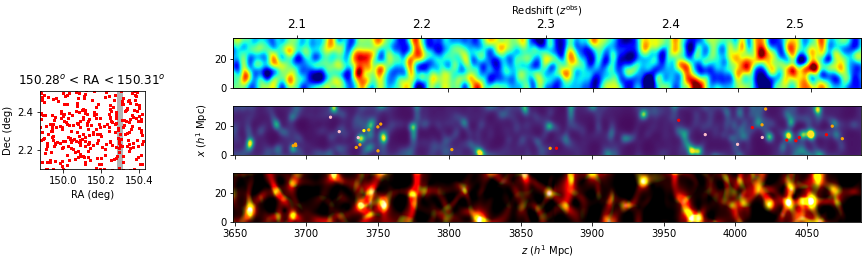}

\includegraphics[width=0.85\textwidth]{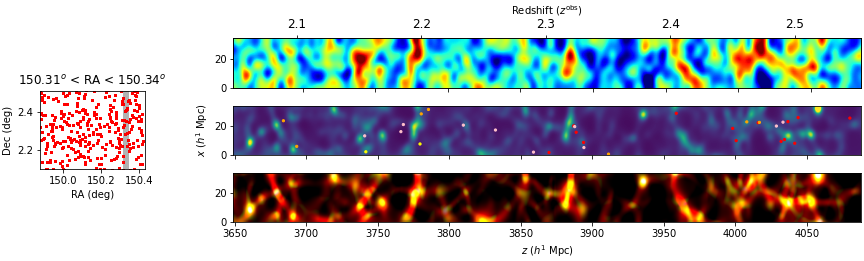}

\caption{\label{fig:c_slicemaps}
Continued from Figure \ref{fig:b_slicemaps}.
}
\end{figure*}

The primary products are the binary files containing the IGM tomographic map, which spans comoving dimensions of 
$30\,\hMpc \times 24\,\hMpc \times 438\,\hMpc$ in the $[x,y,z]$ dimensions, respectively, with binning in units of 0.5\,\hMpc. In addtion we make available our TARDIS reconstructions which cover the same volume but with binning in units of 1.0\,\hMpc. For both products we provide both the direct tomographic
 reconstruction of the data, as well as a version which has been Gaussian-smoothed with a $\sigma=2\,\hMpc$ kernel; the latter
 is the version shown in the visualizations in Figures~\ref{fig:screenshot} and \ref{fig:x3d}. We show slices of our various dataproducts in Figure  \ref{fig:a_slicemaps}, \ref{fig:b_slicemaps} and \ref{fig:c_slicemaps}.

\begin{figure*}\centering

\includegraphics[width=0.85\textwidth]{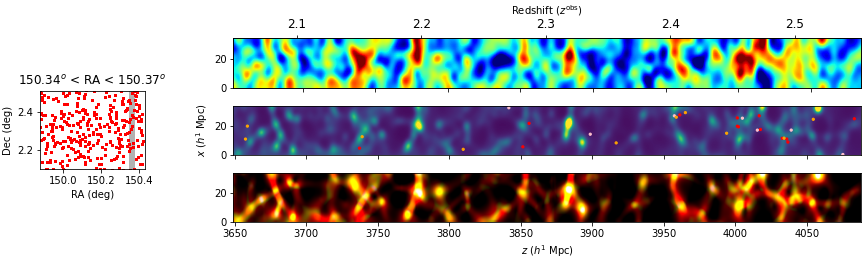}

\includegraphics[width=0.85\textwidth]{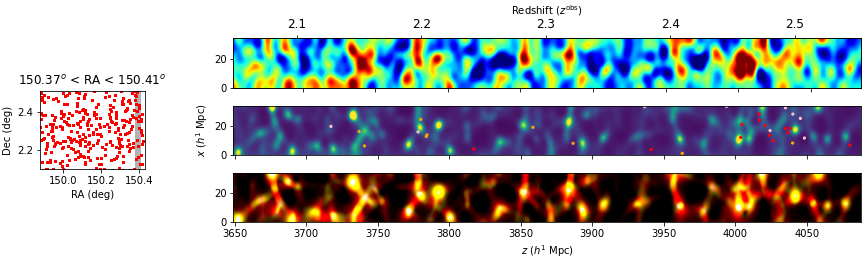}

\caption{\label{fig:d_slicemaps}
Continued from Figure \ref{fig:c_slicemaps}.
}
\end{figure*}

\bibliographystyle{aasjournal}

\bibliography{lyaf_kg,apj-jour,lss_galaxies,my_papers,DR2_new}

\end{document}